\documentclass[12pt]{article}
\usepackage{amsthm,amsmath,amssymb,bm,bbm,graphicx,natbib,color}
\usepackage{enumerate}

\usepackage[ruled]{algorithm2e}
\usepackage{multirow}
\usepackage{float}
\usepackage{url} 
\usepackage[colorlinks=true, citecolor=blue]{hyperref}
\usepackage{multirow,booktabs}

\graphicspath{{figures/}}

\newcommand{\blind}{1}

\usepackage{subcaption}
\usepackage{xcolor}
\usepackage{soul}
\usepackage{booktabs}

\usepackage{mathtools}

\newcommand{\wh}{\widehat}
\newcommand{\wt}{\widetilde}
\newcommand{\rmd}{\mathrm{d}}

\newcommand{\tE}{\text{E}}

\DeclareMathOperator*{\argmin}{arg\,min}

\newcommand{\bc}{\mathbf{c}}
\newcommand{\bd}{\mathbf{d}}

\newcommand{\bh}{\mathbf{h}}

\newcommand{\bbeta}{\bm{\beta}}

\newcommand{\btheta}{\bm{\theta}}
\newcommand{\bgamma}{\bm{\gamma}}

\numberwithin{equation}{section}

\newcommand{\bigcell}[2]{\begin{tabular}{@{}#1@{}}#2\end{tabular}}
\usepackage{tabularx}
\newcolumntype{R}{>{\raggedleft\arraybackslash}X}
\newcolumntype{L}{>{\raggedright\arraybackslash}X}

\begin{document}

\def\spacingset#1{\renewcommand{\baselinestretch}%
{#1}\small\normalsize} \spacingset{1}


\if1\blind
{
  \title{\bf Dynamical Modeling for non-Gaussian Data \\with High-dimensional Sparse Ordinary Differential Equations}
  \author{Muye Nanshan, Nan Zhang
\hspace{.2cm}\\
    School of Data Science, Fudan University\\
    and \\
    Xiaolei Xun \\
    Global Statistics and Data Science, BeiGene\\
    and \\
    Jiguo Cao \\
    Department of Statistics and Actuarial Science, Simon Fraser University}
  \date{}
  \maketitle
} \fi

\if0\blind
{
  \bigskip
  \bigskip
  \bigskip
  \begin{center}
    {\LARGE\bf Dynamical Modeling for non-Gaussian Data \\with High-dimensional Sparse Ordinary Differential Equations}
\end{center}
  \medskip
} \fi

\bigskip
\begin{abstract}
Ordinary differential equations (ODE) have been widely used for modeling dynamical complex systems. For high-dimensional ODE models where the number of differential equations is large, it remains challenging to estimate the ODE parameters and to identify the sparse structure of the ODE models. Most existing methods exploit the least-square based approach and are only applicable to Gaussian observations. However, as discrete data are ubiquitous in applications, it is of practical importance to develop dynamic modeling for non-Gaussian observations. New methods and algorithms are developed for both parameter estimation and sparse structure identification in high-dimensional linear ODE systems. First, the high-dimensional generalized profiling method is proposed as a likelihood-based approach with ODE fidelity and sparsity-inducing regularization, along with efficient computation based on parameter cascading. Second, two versions of the two-step collocation methods are extended to the non-Gaussian set-up by incorporating the iteratively reweighted least squares technique. Simulations show that the profiling procedure has excellent performance in latent process and derivative fitting and ODE parameter estimation, while the two-step collocation approach excels in identifying the sparse structure of the ODE system. The usefulness of the proposed methods is also demonstrated by analyzing three real datasets from Google trends, stock market sectors, and yeast cell cycle studies.
\end{abstract}

\noindent%
\vfill

\newpage
\spacingset{1.5} 

\section{Introduction}

Ordinary differential equations (ODE) are widely used for complex dynamic system modeling in biology, engineering, econometrics, and other scientific and social applications. For example, massive gene expression profiles are available with the advancement of second-generation sequencing technology. Modeling their dynamics using gene regulatory networks has drawn significant interest from both biomedical and statistical research communities \citep{stuart2003gene,yuan2006hidden,hecker2009gene,polynikis2009comparing,lu2011high,wu2014sparse}.
In computational sociology, public opinion sensing and trend analysis have emerged from the advent of the big data revolution \citep{dodds2011temporal,sloan2015tweets}. Massive datasets, such as Google searches or Twitter posts, are collected daily or even hourly, which enables social scientists to extract interesting temporal or spatial patterns via dynamic modeling. The main purpose of this article is to propose new methods and algorithms to estimate the ODE parameters and to identify the sparse structure for high-dimensional ODE models with non-Gaussian observations.

A general first-order ODE system can be described as
\begin{equation}\label{eq:ode-gen}
	\btheta'(t)=f(\btheta(t),\bbeta),
\end{equation}
where the vector $\btheta(t)=(\theta_1(t),\dots,\theta_p(t))^\top$ collects $p$ processes while $\btheta'(t)$ is the first-order derivative of $\btheta(t)$, function $f=(f_1,\dots,f_p)$ describes the dependence between processes and their derivatives, $\bbeta$ is the vector of ODE parameters to be estimated. Typically, the processes are indexed with time $t$ and some initial conditions, for example, $\btheta(0)=\btheta_0$, are assumed for the ODE system \eqref{eq:ode-gen} as well.

In practice, observations from the dynamic system are measured according to the realizations of latent processes $\btheta(t)$ at discrete time points. Estimation of ODE parameters from noisy data remains a challenging problem \citep{ramsay2007parameter, wu2014sparse,hall2014quick,chen2017network, wu2019parameter,dai2021kernel}. In general, parameter estimation procedures fall into three categories. The first approach is based on a data fitting process by nonlinear least squares. Given a set of initial ODE parameters, the ODE solutions are approximated by numerical methods, for example, the Runge-Kutta algorithm. Then the ODE parameters are updated with the nonlinear least squares. This approach is computationally intensive and can be potentially inaccurate due to iterative numerical approximations. 
The second approach is the two-step collocation, where the basis expansions are exploited to approximate the ODE solutions. \citet{varah1982spline} proposed to fit the processes via data smoothing methods, followed by a second stage of minimizing a least-square criterion based on the ODE system to estimate the ODE parameters. 
Because of its computational advantage, two-step collocation gains much popularity in the development of methodology and applications \citep{liang2008parameter,lu2011high,brunel2014parametric,wu2014sparse,dattner2015optimal} and is further improved by iterative principal differential analysis \citep{ramsay1996principal,poyton2006parameter}. However, the performance of two-step procedures relies heavily on the smoothing step, while the amount of roughness regularization is hard to control. 
The third approach is the generalized profiling procedure \citep{ramsay2007parameter}, which also represents ODE solutions with basis expansion as with two-step collocation methods. The essential difference is the inclusion of an ODE-induced penalty that controls the fidelity of the processes to the ODE system. The basis coefficients and ODE parameters are then estimated simultaneously from a penalized criterion using the parameter cascading algorithm \citep{cao2007parameter}. From a theoretical perspective, \citet{qi2010asymptotic} derived an upper bound on the uniform norm of the difference between the true underlying solutions and their approximations, and proved the consistency and asymptotic normality of the estimation procedure.

More recently, there has been growing interest in high-dimensional ODE systems where the number of processes $p$ is large. For instance, the high-dimensional time-course gene expression data enables biomedical researchers to model the regulatory behaviors via a large-scale directed graphical network model. Such a task is called network recovery. The ODE system \eqref{eq:ode-gen} naturally serves for this purpose by relating the dynamics of each process with all the processes in the system, and a sparse network structure can be further imposed.
\citet{lu2011high} considered the high-dimensional linear ODE for dynamic gene regulatory network identification and applied the smoothly clipped absolute deviation \citep{fan2001variable} approach for variable selection.
\citet{wu2014sparse} further relaxed the linear assumption and investigated a sparse additive ODE model using a two-stage procedure coupled with the adaptive group Lasso technique \citep{wang2008note} to deal with nonlinear effects.
\citet{chen2017network} proposed an efficient procedure using the integrated form of the ODE to bypass numerical difficulty in the derivative estimation and adopted the group Lasso \citep{yuan2006model} for variable selection. \citet{wu2019parameter} recently developed a matrix factorization based approach to ultra-high dimensional linear ODE models for parameter estimation and variable selection. To our best knowledge, existing procedures for high-dimensional ODE models are two-stage approaches. 

Besides, most of the existing work assumes that observations of the ODE system are contaminated with Gaussian noises. Therefore, least-squares estimation is conveniently adopted. However, non-Gaussian observations are commonly encountered in real applications, for example, short read count data from RNA sequencing \citep{nagalakshmi2008transcriptional}, bisulfite sequencing data for DNA methylation analysis \citep{cokus2008shotgun}, and direction of change in the stock price over time \citep{huang2005forecasting}. 
The literature on non-Gaussian data analysis with the ODE system is rare. \citet{miao2014generalized} developed a likelihood-based parameter estimation and inference for generalized ODE models. Its extension to high-dimensional ODE models, however, is still unknown.

Motivated by network recovery tasks for time-course non-Gaussian data, this paper focuses on the parameter estimation and sparse structure identification for high-dimensional linear ODE systems with a likelihood-based approach. To facilitate versatile analysis of non-Gaussian data, we assume the observations follow a distribution from the exponential family, where $\theta_j(t)$ is known as the canonical parameter in the context of generalized linear models \citep{mccullagh1989generalized,wood2017generalized}.
Assume that $t\in[0,1]$ without loss of generality. Given a set of discrete time points $t_1,\dots, t_n$, denote by $y_{ij}$ the measurement according to the $j$th latent process $\theta_j(t)$ at time $t=t_i$, $j=1,\dots,p$. Then, the conditional distribution of $y_{ij}$ given $\theta_j(t_i)$ admits a density function as
\begin{equation*}
	f(y_{ij} \mid \theta_j(t_i))=\exp\left\{\frac{y_{ij}\theta_j(t_i)-b(\theta_j(t_i))}{a(\phi)} + c(y_{ij},\phi)\right\},
\end{equation*}
where $a>0, b, c$ are known functions, $\phi$ is either known or considered as a nuisance parameter. 
Let $(y_{1j},\dots,y_{nj})^\top$ be the vector of observations from the latent process $\theta_j(t)$, and correspondingly the canonical parameter vector be $(\theta_j(t_1),\dots,\theta_j(t_n))^\top$.
Imposing a linear structure on the general model \eqref{eq:ode-gen}, we investigate in this work the modeling of the dynamics among latent processes $\{\theta_j(t): j=1,\dots,p\}$ with a high-dimensional linear ODE system, that is
\begin{equation}\label{eq:ode-linear}
	\theta_j^{'}(t)=\gamma_{j0} + \sum_{k=1}^p \gamma_{jk} \theta_k(t), \qquad j=1,\dots,p.
\end{equation}

In this article, we develop new methods and algorithms for both parameter estimation and sparse structure identification in high-dimensional linear ODE systems. First, we propose the high-dimensional generalized profiling method along with a computationally efficient procedure based on parameter cascading \citep{ramsay2007parameter,cao2007parameter}. It solves a hierarchical optimization for parameter estimation and variable selection: an outer optimization concerning the ODE parameters under sparsity regularization is performed subject to an inner optimization where latent processes expanded with basis functions are fitted by minimizing a weighted sum of data fitting and ODE fidelity criteria given ODE parameters. In particular, we regularize the structural ODE parameters based on individual differential equation and mitigate the computational burden for parameter estimation in the high-dimensional ODE system. Moreover, there are two tuning parameters involved in our procedure: one controls the balance between data fitting and ODE fidelity in the inner optimization while the other regularizes the sparsity or model complexity in the outer optimization. Their interaction may affect the overall convergence performance of the procedure in a complicated way. Due to the non-convexity nature of our objective function, we carefully design the tuning and stopping rules according to the performance of parameter estimation to help escape local minima \citep{carey2021fast}. The global convergence of the proposed algorithm is analyzed. 
Next, we extend the two-step collocation methods \citep{wu2014sparse,chen2017network}, which are recently proposed for high-dimensional ODE models with Gaussian observations, to the non-Gaussian set-up. Two versions, corresponding to the vanilla collocation \citep{varah1982spline} and the graph reconstruction via additive differential equations (GRADE) \citep{chen2017network}, are developed under the likelihood-based framework. Efficient computation is feasible by applying the iteratively reweighted least squares technique \citep{wood2017generalized}.
Finally, we apply the proposed methods to simulated and real data sets. In general, the profiling method is more efficient than two-step collocation methods in estimating the latent processes, their derivatives, and the structural ODE parameters, while one two-step collocation method excels in identifying the sparse structure of the ODE system. To sum up, the proposed methods present a versatile toolbox for parameter estimation and sparse structure identification in high-dimensional linear ODE systems.

The remainder of the article is organized as follows. Our profiled estimation approach is developed in Section~\ref{sec:method}. Detailed computational procedure and its global convergence are discussed in Section~\ref{sec:compt}.
In Section~\ref{sec:two-step}, we extend two-step collocation methods to model non-Gaussian observations. Section~\ref{sec:simu} compares empirical performance of the proposed methods. We analyze three real data examples in Section~\ref{sec:app} with dynamical modeling approaches. Section~\ref{sec:conclude} concludes the article and Appendix collects some technical details.

\section{High-dimensional Generalized Profiling}\label{sec:method}

This section introduces the proposed approach for simultaneous parameter estimation and sparse structure identification in a high-dimensional linear ODE model for non-Gaussian data under the penalized likelihood estimation framework.

Denote by $\Gamma = (\bgamma_1, \dots, \bgamma_p)$ the parameter matrix of the ODE model~\eqref{eq:ode-linear}, where $\bgamma_j=(\gamma_{j0},\dots, \gamma_{jp})^\top\in\mathbb{R}^{p+1}$, for $j=1,\dots,p$.
These ODE parameters $\Gamma$ are of primary interest in order to understand the network structure, called structural parameters hereafter. On the other hand, the latent processes $\theta_j$'s are treated as nuisance parameters. Denote by $y_{ij}$ and $\theta_j(t_i)$ the observation and the canonical parameter of the $j$th latent process at time $t_i$, respectively. Under the profiling scheme \citep{ramsay2007parameter}, an intermediate fit of latent processes $\wh\btheta(t;\Gamma)=(\wh\theta_1(t;\Gamma),\dots, \wh\theta_p(t;\Gamma))$ minimizes the following penalized likelihood criterion,
\begin{equation}\label{eq:theta_gp}
	-\frac{1}{np}\sum_{i=1}^n\sum_{j=1}^p\left\{y_{ij}{\theta_j(t_i)}-b(\theta_j(t_i))\right\}
	+ \lambda_\theta \sum_{j=1}^p\int_0^1 \left\{\theta_j^{'}(t)-\gamma_{j0}-\sum_{k=1}^p \gamma_{jk} \theta_k(t)\right\}^2 \,\rmd t,
\end{equation}
where the likelihood part measures fidelity to data, the ODE fidelity part measures the extent to which latent processes fail to satisfy the ODE system, and the tuning parameter $\lambda_\theta$ controls the amount of regularization.
Furthermore, with $\wh\btheta(t;\Gamma)$ plugged in, an estimate of the structural parameters can be obtained by minimizing a data fitting criterion with respect to $\Gamma$,
\begin{equation}\label{eq:gamma_gp}
	-\frac{1}{np}\sum_{i=1}^n\sum_{j=1}^p\{y_{ij}{\wh\theta_j(t_i; \Gamma)}-b(\wh\theta_j(t_i; \Gamma))\}.
\end{equation}
The generalized profiling procedure proceed iteratively with a non-decreasing sequence of $\lambda_\theta$ under certain rules such that the fitted processes adhere to the ODE. Identifiable issue and asymptotic behavior of the estimation procedure are studied by \citet{ramsay2007parameter} and \citet{qi2010asymptotic}.

Although the generalized profiling method provides a computationally efficient treatment for the challenging ODE parameter estimation, it can only handle relatively small-scale models \citep{wu2019parameter}. On the one hand, for a $p$-dimensional linear ODE system, we have $p^2+p$ ODE parameters to estimate in \eqref{eq:gamma_gp}. If we further approximate the latent process $\theta_j(t)$ by basis expansion $\bc_j^\top \bh_j(t)$, where $\bh_j(t)$ is an $m_j$-dimensional basis vector and $\bc_j$ is the coefficient vector, then \eqref{eq:theta_gp} becomes
\begin{align*}\label{eq:theta_gp_expansion}
	-\frac{1}{np}\sum_{i=1}^n\sum_{j=1}^p&\left\{y_{ij}{\bc_j^\top \bh_j(t_i)}-b(\bc_j^\top \bh_j(t_i))\right\} \\
	&\qquad\qquad + \lambda_\theta \sum_{j=1}^p\int_0^1 \left\{\bc_j^\top \bh_j^{'}(t)-\gamma_{j0}-\sum_{k=1}^p \gamma_{jk} \bc_j^\top \bh_k(t)\right\}^2 \,\rmd t,
\end{align*}
and the total number of nuisance parameters $\sum_{j=1}^p m_j$ can be huge. Therefore, a direct application of the standard generalized profiling procedure to parameter estimation for high-dimensional linear ODE is computationally demanding.
On the other hand, the structural parameters obtained from \eqref{eq:gamma_gp} indeed infer an interaction network among the latent processes, in the sense that a nonzero $\gamma_{jk}$ implies that $\theta_k(t)$ has an effect on the change of $\theta_j(t)$. For better interpretation and to avoid potential over-fitting, it is reasonable to introduce some sparsity for the structural parameters. For example, the Lasso and its variants \citep{tibshirani1996regression, yuan2006model, zou2006adaptive}, the smoothly clipped absolute deviation  (SCAD) \citep{fan2001variable} and the minimax concave penalty (MCP) \citep{zhang2010nearly} have been extensively studied and used to recover probabilistic graphical structures \citep{yuan2007model,fan2009,voorman2014graph}.

To address the above computational issues, we first notice that the data fidelity term in the penalized criterion \eqref{eq:theta_gp} can be decomposed into sums of the likelihood for $p$ individual processes. Meanwhile, the penalty term, being a squared $L_2$ norm of differential equations, does not admit such decomposable property. Therefore, we propose to regularize the estimate of $\theta_j$ only by the corresponding $j$th differential equation. Specifically, when estimating $\theta_j$ given other $\{\theta_k: k\neq j\}$ at their most recent updates, we obtain $\wh\theta_j(t; \bgamma_j)$ by minimizing
	\begin{equation}\label{eq:G}
		G_j(\theta_j; \bgamma_j)=
		-\frac{1}{n}\sum_{i=1}^n\{y_{ij}\theta_j(t_i)-b(\theta_j(t_i))\}
		+ \lambda_{\theta,j} \int_0^1 
		\biggl\{ \theta_j^{'}(t) - \gamma_{j0} - \sum_{k=1}^p \gamma_{jk}\theta_k(t)\biggr\}^2  \,\mathrm{d}t,
	\end{equation}
for $j=1,\dots,p$. For simplicity, we use the same tuning parameter for individual sub-problems, that is $\lambda_{\theta,j}=\lambda_\theta$ for $j=1,\dots, p$. Optimizing $G_j$ involves only $p+1$ structural parameters in the vector $\bgamma_j$ and hence the computational complexity is greatly reduced. The benefit of using \eqref{eq:G} is justified from two aspects. First, it is computationally infeasible to estimate a large number of ODE parameters jointly by directly applying the original generalized profiling criterion \eqref{eq:theta_gp} to the high-dimensional ODE system. Our new formulation decouples the dependency of $\wh\btheta(t;\Gamma)$ on the matrix $\Gamma$ into individual dependencies of $\theta_j(t;\bgamma_j)$ on the vector $\bgamma_j$. Second, from the perspective of penalized estimation, it improves the estimation for the latent process and the ODE structural parameters by employing differential equations to regularize data smoothing. 

We remark on the potential risk of employing \eqref{eq:G} instead of \eqref{eq:theta_gp} when estimating the latent processes. Note that \eqref{eq:theta_gp} aggregates all the differential equations to update the latent processes altogether such that the estimates will follow the ODE system jointly. In contrast, our method uses a single differential equation to regularize the estimation of each latent process. When the tuning parameter $\lambda_\theta$ increases, the parallel updating procedure \eqref{eq:G} over $j=1,\dots,p,$ is expected to achieve an approximation in a marginal way to the joint estimation by \eqref{eq:theta_gp}. The simulation example introduced in Section S1 of the Supplementary Material shows that the approximation by \eqref{eq:G} performs reasonably well, although the joint method \eqref{eq:theta_gp} has a more accurate estimate for ODE parameters.

Next, to induce sparsity to the structural parameter matrix, we estimate $\bgamma_j$ by minimizing
\begin{equation}\label{eq:H}
	H_j(\bgamma_j)=-\frac{1}{n}\sum_{i=1}^n\{y_{ij}{\wh\theta_j(t_i; \bgamma_j)}-b(\wh\theta_j(t_i; \bgamma_j))\}
	+  \mathrm{PEN}_{\lambda_{\gamma,j}}(\bgamma_{j}),
\end{equation}
where the penalty function $\mathrm{PEN}_{\lambda_{\gamma,j}}(\bgamma_{j})$ with tuning parameter $\lambda_{\gamma,j}>0$ induces sparsity for the structural parameter of the $j$th differential equation. Here we also assume for simplicity that $\lambda_{\gamma,j}=\lambda_\gamma$ for $j=1,\dots,p$. If the fitted structural parameter vector $\wh\bgamma_j$ is zero, then we say other latent processes have no impact on $\theta_j(t)$. Any zero element in $\wh\bgamma_j$ implies that the corresponding process has no influence on $\theta_j(t)$. The amount of sparsity regularization is typically determined by Bayesian information criterion (BIC) type principles, which have been adopted in other ODE parameter estimation approaches \citep{wu2014sparse,chen2017network}.

Our new profiling estimation procedure for high-dimensional linear ODE systems consists of two objective functions 
\eqref{eq:G} and \eqref{eq:H}, which are referred to as inner and outer criteria, respectively. Such a multi-criterion optimization problem is challenging due to non-convexity and non-differentiability. Specifically, we approximate the latent processes with basis expansion in the inner optimization, and basis coefficients can be solved efficiently with the Newton-Raphson method. However, the dependence of $\wh\theta_j(t; \bgamma_j)$ on $\bgamma_j$ is complicated and in general non-linear, which leads to the non-convexity of $H_j$.
Moreover, the sparsity-inducing penalty in $H_j$ is non-differentiable at zero, making the Gauss-Newton scheme adopted by \citet{ramsay2007parameter} invalid under this scenario.

Recent advances of derivative-free optimization algorithms \citep{powell2006newuoa,zhang2010derivative} may provide a viable solution. Nevertheless, they are in spirit joint optimization algorithms designed for general purpose and are thus not tailored for our specific problem. In contrast, our profiling procedure enjoys not only estimation efficiency but also algorithmic efficiency due to the use of analytical expressions of derivatives. Computational details are presented in the next section. In brief, after obtaining an estimate $\wh\theta_j(t; \bgamma_j)$ given the structural parameters, we linearize the likelihood component in \eqref{eq:H} and formulate the outer optimization as a parameter estimation problem for a penalized generalized linear model. Therefore, the structural parameters can be readily updated by the iterative reweighted least-squares (IRLS). Through an iterative scheme between inner and outer optimizations, our profiling procedure provides ODE parameter estimates and latent process fits and identifies the sparse structure of the ODE model.

\section{Computation}\label{sec:compt}

In this section, we provide computational details of our profiling procedure for high-dimensional linear ODE and analyze its global convergence. Minimizing criteria in \eqref{eq:G} and \eqref{eq:H} are referred as inner and outer optimizations. The structural parameters $\Gamma=(\bgamma_1, \dots, \bgamma_p)$ is of our primary interest, while the latent process fits by the inner optimization is regarded as a nuisance parameter. In our profiling scheme, whenever $\bgamma_j$ changes by minimizing $H_j$ in the outer, latent process fits are then updated by solving the inner criterion $G_j$. Details are provided in Algorithm~\ref{algo:hdgp}. In addition, two tuning parameters are involved in the profiling procedure, and their complex interaction affects the overall algorithmic performance because of the non-convexity of the optimization. In the following, we split the discussion into inner and outer parts. Then we discuss the practical strategy of tuning parameter selection and the global convergence of the proposed algorithm.

\begin{algorithm}[!hb]
	\SetAlgoLined
	\DontPrintSemicolon
	\BlankLine
	\KwIn{Observations $\{y_{ij}: i=1,\dots,n; j=1,\dots, p\}$, initial ODE paramters $\Gamma^{(0)}=(\bgamma_1^{(0)},\dots, \bgamma_p^{(0)})$, and fixed tuning parameters $\lambda_\theta$ and $\lambda_\gamma$.}
	
	\KwOut{Estimated ODE parameters $\wh\Gamma=(\wh\bgamma_1,\dots, \wh\bgamma_p)$.}
	\BlankLine
	\Repeat{Estimated ODE parameter $\wh\Gamma$ converges.}{
		At step $s\geq 1$, the current estimate is $\wh\Gamma^{(s)}=(\wh\bgamma_1^{(s)},\dots,\wh\bgamma_p^{(s)})$.
		
		\For{$1\leq j\leq p$}{
			Update $\bgamma_j$ via the profiling procedure.
			
			\Repeat{$\wt\bgamma_j$ converges, and set $\bgamma_j^{(s+1)}= \wt\bgamma_j$.}{
				1. Given current $\wt\bgamma_j$, obtain the basis coefficient estimate $\bc_j^*(\wt\bgamma_j)$ for \\the $j$th latent process  in the inner optimization.
                
                2. Apply basis expansion and update $\bgamma_j$ via minimizing the penalized \\ reweighted least squares.
			}
		}
	}
	\caption{High-dimensional linear ODE for non-Gaussian data}
	\label{algo:hdgp}
\end{algorithm}

\subsection{Inner Optimization} \label{sec:inner}

The inner procedure aims at finding an accurate estimate for latent processes given the structural parameter $\Gamma$. Similar to the two-step collocation method \citep{varah1982spline} and the generalized profiling \citep{ramsay2007parameter}, we represent latent processes by basis expansion. Suppose $\bh_j(t) = (\phi_{j1}(t),\ldots,\phi_{jm_j}(t))$ is a set of basis functions for the $j$th process such that $\theta_j(t) = \bc_j^\top \bh_j(t).$ Choices of basis functions include polynomials, truncated power functions and splines. In our numerical study, we use B-spline due to its numerical stability and excellent empirical performance. For notation simplicity, we use the same basis $\bh(t)$ for all latent processes. 

Although critical in optimization, the basis coefficients $\bc_j$, $j=1,\dots,p$, are often not of direct concern and thus considered as nuisance parameters. Observing that $G_j(\theta_j;\bgamma_{j})$ is convex with respect to the basis coefficients $\bc_j$, we can apply the Newton-Raphson scheme directly. When the Hessian of $G_j(\theta_j;\bgamma_{j})$ is invertible, we can start with an initial guess of $\bc_j$ and iteratively obtain
$$
\bc_j^{(r+1)} = \bc_j^{(r)} 
- \biggl(\frac{\partial^2 G_j}{\partial \bc_j \partial \bc_j^\top}\bigg|_{\bc_j^{(r)}} \biggr)^{-1} 
\biggl(\frac{\partial G_j}{\partial \bc_j}\bigg|_{\bc_j^{(r)}}\biggr), \quad r\geq 1.
$$
Analytical expressions of the derivatives involved in the above updating rule are given in \ref{sec:derivative}.

\subsection{Outer Optimization}\label{sec:outer}

The outer optimization is designed for updating $\bgamma_j$ with a regularized likelihood objective function \eqref{eq:H}. Denote by $\bc_j^*(\bgamma_j)$ the optimal basis coefficients for $\theta_j(t;\bgamma_j)$ obtained from the inner optimization given the current $\bgamma_j$.  Observing that the dependence of $\bc_j^*(\bgamma_j)$ on $\bgamma_j$ is implicit and possibly complicated, we propose to linearize the likelihood component in \eqref{eq:H} and transform the optimization to finding the maximum likelihood estimate of a generalized linear model. The solution can then be readily obtained by the iteratively reweighted least squares (IRLS), see \cite{wood2017generalized} for more detail.

Let $\wt\bgamma_j$ be the most recent update of $\bgamma_j$. First, we linearize the $\bc_j^*(\bgamma_j)$ at $\wt\bgamma_j$ which,
\begin{equation}\label{eq:linearization}
	\bc_j^*(\bgamma_j) \approx \bc_j^*(\wt\bgamma_j) + \frac{\partial \bc_j^*(\bgamma_j)}{\partial\bgamma_j^\top}\bigg|_{\wt\bgamma_j} (\bgamma_j - \wt\bgamma_j),
\end{equation}
where the derivative $\partial \bc_j^*/\partial \bgamma_j$ is explicitly derived using the implicit function theorem in \ref{sec:derivative}.
Hence, $\wh{\theta}_j(t; \bgamma_j)$ in its basis expansion form can be approximated by a linear function of $\bgamma_j$. As a result, the outer objective function \eqref{eq:H} now becomes a penalized likelihood function of a generalized linear model.
Second, we apply the IRLS and update our estimate of $\bgamma_j$. Let  $\wt\theta_j(t)=\wh\theta_j(t;\wt\bgamma_j)$ be latent process fit given the structural parameter $\wt\bgamma_j$. Based on the theory of generalized linear models,
the observation $Y_j$ according to the latent process $\wt\theta_j(t)$ admits properties of $\tE(Y_j|\wt\theta_j(t))=b^{'}(\wt\theta_j(t))=\wt\mu_j(t)$ and $\text{var}(Y_j|\wt\theta_j(t))=b^{''}(\wt\theta_j(t))a(\phi)=\wt{v}_j(t)a(\phi)$, where functions $a,b$ and parameter $\phi$ follow from the exponential family specification. Write $\wt{u}_{ij} = -y_{ij}+b^{'}(\wt\theta_j(t_i))=-y_{ij}+ \wt\mu_j(t_i)$ and $\wt{w}_{ij}=b^{''}(\wt\theta_j(t_i))=\wt v_j(t_i)$.
The IRLS algorithm applies a quadratic approximation to the log-likelihood, that is, at $\theta_j = \wt\theta_j$,
\[
-y_{ij}\wh \theta_j(t_i; \bgamma_j)+b(\wh \theta_j(t_i; \bgamma_j))
\approx \frac{1}{2} \wt{w}_{ij} \left\{\wt{y}_{ij} - \wh \theta_j(t_i; \bgamma_j)\right\}^2 + C_{ij},
\]
where 
$\wt{y}_{ij}=\wt\theta_j(t_i)-\wt{u}_{ij} /\wt{w}_{ij}
$ and $C_{ij}$ is independent of $\wt \theta_j(t_i)$. In conjunction with the linear approximation of $\theta_j(t_i; \bgamma_j)$, it amounts to solving a penalized linear least squares to update the estimate for structural parameter $\bgamma_j$. Efficient algorithms are available for different sparsity penalty choices $\textrm{PEN}(\cdot)$.

\subsection{Tuning Parameter Selection}\label{sec:para_select}

There are two tuning parameters involved in our profiling procedure, which jointly affect the algorithmic performance. On the one hand, $\lambda_\theta$ in the inner optimization controls the amount of regularization regarding the differential equations. We define the aggregated ODE fidelity criterion as
\begin{equation}\label{eq:ode-fidelity}
	\sum_{j=1}^p \int_0^1 \left\{\theta_j^{'}(t)-\gamma_{j0}-\sum_{k=1}^p \gamma_{jk} \theta_k(t)\right\}^2 \,\rmd t.
\end{equation}
Small $\lambda_\theta$ makes optimizing $H_j(\bgamma_j)$ with respect to $\bgamma_j$ more robust to initial guesses, but yields bad approximations to ODE solutions. Large $\lambda_\theta$ gives rise to a difficult optimization problem where $H_j(\bgamma_j)$ is usually not convex and can have many local optima \citep{ramsay2007parameter,qi2010asymptotic,carey2021fast}. On the other hand, $\lambda_\gamma$ in the outer optimization induces a sparse network structure for latent processes with better interpretation, and existing methods such as information criteria can be adopted for tuning. Based on the above discussion, we propose to fix $\lambda_\gamma$ in the outer optimization first, iteratively select a proper $\lambda_\theta$ in the inner optimization, and then determine the best $\lambda_\gamma$ via the Bayesian information criterion. In detail, suppose we choose $\lambda_\gamma$ from a sequence of candidate values. Then, we initialize $\lambda_\theta$ with a small value and moderately increase it via an iterative scheme. At each iteration, $\wh\btheta$ and $\wh\Gamma$ are repeatedly estimated for the current $\lambda_\theta$, which are then used as initial values in the fitting procedure with the next larger $\lambda_\theta$. The iterative scheme stops when the estimated ODE parameters converge, and thus $\lambda_\theta$ is decided. The change of the estimated ODE parameters should be small when there is only a moderate increase in $\lambda_\theta$. Therefore, with a conservatively increasing sequence of $\lambda_\theta$, every estimated $\wh\Gamma$ is much likely to be a proper initialization for the next iteration. Details of the iterative selecting scheme for $\lambda_\theta$ given a fixed $\lambda_\gamma$ are as follows. 
\begin{enumerate}[(1)]
	\item Start with a small positive $\lambda_\theta^{(0)}$. Choose $\Delta^{(0)}$ as an initial incremental factor.  
	\item At the $u$th iteration where $u\geq 0$, obtain the fitted latent processes $\wh\btheta^{(u)}$ and $\wh\Gamma^{(u)}$ via our profiling procedure, and evaluate the ODE fidelity \eqref{eq:ode-fidelity} based on the estimates.
	\begin{enumerate}
		\item If the absolute percentage of change in the ODE fidelity \eqref{eq:ode-fidelity} is below a threshold constant, then we update $\lambda_\theta^{(u+1)}=\lambda_\theta^{(u)}\times \Delta^{(u)}$. 
		\item Otherwise, we need to downsize the incremental factor, for example, set $\Delta^{(u)}=\Delta^{(u-1)}/2$, which ensures that the ODE fidelity \eqref{eq:ode-fidelity} varies little among iterations.
	\end{enumerate}
	\item When the successive ODE parameter estimates are closed enough, we stop iteration; otherwise, repeat previous steps.
\end{enumerate}

Our iterative tuning strategy treats $\lambda_\theta$ as a function of $\lambda_\gamma$. Hence, after $\lambda_\theta$ is selected for each fixed $\lambda_\gamma$ from a sequence of candidate values, we can evaluate the following BIC and choose the best $\lambda_\gamma$,
\begin{equation*}
	\mathrm{BIC}(\lambda_\gamma) 
	=    -\frac{1}{np}\sum_{i=1}^n\sum_{j=1}^p\left\{y_{ij}{\wh\theta_j(t_i;\lambda_\gamma)}-b(\theta_j(t_i;\lambda_\gamma))\right\}+ k(\mathbf{\lambda}) \log(n),
\end{equation*}
where $\wh\theta_j(t;\lambda_\gamma)$ emphasizes the dependence on $\lambda_\gamma$, and $k(\lambda_\gamma)$ denotes the number of non-zero elements in the resultant ODE parameter $\wh\Gamma(\lambda_\gamma)$.

\subsection{Global Convergence}

Suppose that the estimated latent process $\wh\theta_j(t;\bgamma_j)$ from the inner optimization is a smooth function of $\bgamma_j$, where $j=1,\dots,p$. Let $H(\Gamma) = \sum_{j=1}^p H_j(\bgamma_j)$ be the objective function in the outer optimization for a given tuning parameter $\lambda_\gamma$, where $\Gamma = (\bgamma_1, \dots, \bgamma_p)$. Algorithm~\ref{algo:hdgp} is essentially a block coordinate descent method because it minimizes $H(\Gamma)$ by iteratively updating $\bgamma_j$.
Write $ H_j(\bgamma_j) = \ell_j(\bgamma_j) + \textrm{PEN}_{\lambda_\gamma} (\bgamma_j)$, where $\ell_j(\bgamma_j)$ is the likelihood term and $\textrm{PEN}_{\lambda_\gamma} (\bgamma_j)$ is assumed to be convex. As described in Section~\ref{sec:outer}, the outer optimization is equivalent to updating $\bgamma_j$ to $\bgamma_j + \bd_{j}(\bgamma_j)$, where the descent direction $\bd_{j}(\bgamma_j)$ is the solution to
\begin{equation*}
  \min_{\bd} \nabla \ell_j(\bgamma_j)^\top \bd + \frac{1}{2}\bd^\top Q_j( \bgamma_j) \bd + \textrm{PEN}_{\lambda_\gamma} (\bgamma_j + \bd),
\end{equation*}
where $\nabla \ell_j(\bgamma_j)$ is the gradient of $\ell_j(\bgamma_j)$ and
$$
Q_j( \bgamma_j)
=\frac{1}{n}\sum_{i=1}^n \left\{
b''(\wh\theta_j(t_i; \bgamma_j)) \frac{\partial{\wh\theta_j(t_i; \bgamma_j)}}{\partial \bgamma_j} \left( \frac{\partial{\wh\theta_j(t_i; \bgamma_j)}}{\partial \bgamma_j} \right)^\top\right\}
$$
is a positive definite matrix approximating the Hessian $\nabla^2 \ell_j(\bgamma_j)$.

We follow \citet{tseng2009coordinate} to establish the global convergence. Because the actual value of the Hessian $\nabla^2 \ell_j(\bgamma_j)$ is identical to its expected value under canonical links \citep{mccullagh1989generalized}, the IRLS method described in Section~\ref{sec:outer} remains the same when the Hessian is replaced by the expected Hessian. Then it follows from Lemma~S1 in the Supplementary Material that
\begin{equation}\label{eq:descent}
H_j(\bgamma_j + \bd_{j}(\bgamma_j)) - H_j(\bgamma_j)  \leq 
-\bd_{j}^\top(\bgamma_j) \left[Q_j(\bgamma_j)-\frac{1}{2} \tE\{\nabla^2 \ell_j(\bgamma_j)\} \right] \bd_{j}(\bgamma_j) + o(\|\bd_{j}(\bgamma_j)\|^2).  
\end{equation}
Some algebra yields that
$$
Q_j(\bgamma_j)-\frac{1}{2} \tE\{\nabla^2 \ell_j(\bgamma_j)\}
=\frac{1}{2}Q_j(\bgamma_j) + \frac{1}{2n}\sum_{i=1}^n \left\{ b'(\theta_j^*(t_i)) - b'(\wh\theta_j(t_i, \bgamma_j)) \right\}\, \frac{\partial^2{\wh\theta_j(t_i, \bgamma_j)}}{\partial \bgamma_j \partial \bgamma_j^\top},
$$
where $\theta_j^*(t)$ is the true latent process. The above matrix is positive definite because $b'(\cdot)$ is continuous, provided that $\wh\theta_j(t_i, \bgamma_j)$ is sufficiently close to the truth. It follows from \eqref{eq:descent} that $H_j(\bgamma_j)$ decreases along the iterations and will eventually converge because it is lower-bounded. Moreover, the sequence of descent directions converges to zero due to \eqref{eq:descent}. According to Theorem 1(e) and Lemma 2 of \citet{tseng2009coordinate}, every cluster point of the iterative estimates by Algorithm~\ref{algo:hdgp} exhibits exact zero descent direction, which implies it is indeed a stationary point of $H(\Gamma)$.

Finally, we remark that the above analysis cannot be directly applied to a non-convex $\textrm{PEN}_{\lambda_\gamma} (\cdot)$ such as the SCAD penalty. However, the non-convex penalty can be numerically approximated by local linear or quadratic functions \citep{fan2020statistical}. We would anticipate a similar convergence result but with more involved technical details, which is not pursued in this paper.

\section{Two-step Collocation Methods for non-Gaussian Data}\label{sec:two-step}

Collocation methods have been exploited for both parameter estimation and network reconstruction for various ODE models. In this section, we extend the popular two-step collocation method for high-dimensional linear ODE with non-Gaussian observations. In the large literature on collocation, \citet{varah1982spline, ramsay2007parameter, dattner2015optimal}, and  \citet{wu2019parameter} consider the linear case while recently the nonparametric additive structure is investigated by \citet{henderson2014network, wu2014sparse} and \citet{chen2017network}. Most existing methods are proposed for Gaussian observations and adopt the least square loss function for estimation. In the following, we present two versions of the two-step collocation method for high-dimensional ODE models with non-Gaussian observations: the vanilla collocation based on \citet{varah1982spline} and an extension from graph reconstruction via additive differential equations (GRADE) by \citet{chen2017network}. 

The vanilla two-step method first fits smoothing estimates $\wh\btheta(t)$ to the latent processes with maximum likelihood estimation, and then obtain the structural parameter $\bgamma$ with the estimated processes and their derivatives plugged in. The procedure solves the following optimization problems,
\begin{equation}\label{eq:varah_obj_lasso}
	\wh\bgamma_j = \argmin_{\gamma_{j0}, \bgamma_{j}} \int_0^1 \left| \frac{\mathrm{d}\wh\theta_j(t)}{\mathrm{d}t} - \gamma_{j0} - \sum_{k=1}^p\gamma_{jk} \wh\theta_k(t) \right|^2 \mathrm{d}t + \textrm{PEN}_{\lambda_\gamma}(\bgamma_{j}),
\end{equation}
with
\begin{equation}\label{eq:smoothingsplines}
	\wh\theta_j(t) = \argmin_{\theta \in \mathcal H} -\frac{1}{n}\sum_{i=1}^n\{y_{ij}\theta(t_i)-b(\theta(t_i))\}, \quad 1\leq j\leq p,
\end{equation}
where $\mathcal H$ is a proper reproducing kernel Hilbert space, and the exponential family smoothing splines can be adopted \citep{wahba1995,gu2013smoothing,ma2017adaptive}.
The performance of the vanilla two-step collocation method relies on the estimation accuracy of $\wh\theta_j(t)$ and its derivatives. Although statistical convergence has been established, it is in practice hard to tune the smoothing procedure to achieve the optimality \citep{liang2008parameter,brunel2014parametric}.

Another extension is based on the GRADE method \citep{chen2017network}. It avoids the derivative estimation issue in the vanilla collocation method, and instead considers the ODE fidelity term in its integral form. Similar to the vanilla two-step method, the GRADE method first obtains the smoothing estimates of latent processes from observations as in \eqref{eq:smoothingsplines}.
Using integrated basis functions $\wh\Theta_j(t)= \int_0^t \wh\theta_{j}(t)\, \mathrm{d}t, j=1,\dots,p$, one can express
$$\wt\theta_j(t) = C_{j0} + \gamma_{j0}\,t + \sum_{k=1}^p \gamma_{jk} \wh\Theta_k(t),$$ according to the integrated differential equations.
Finally, we solve the following optimization problems to obtain 
\begin{equation}\label{eq:GRADE-basis}
	\wh\bgamma_j = \argmin_{C_{j0}, \gamma_{j0}, \bgamma_j} \frac{1}{n}\sum_{i=1}^n \left\{y_{ij}\,\wt\theta_j(t_i) - b(\wt\theta_j(t_i))\right\} + \textrm{PEN}_{\lambda_\gamma}(\bgamma_{j}).
\end{equation}
The GRADE method is initially developed for nonparametric additive ODE models and naturally adapts to the linear case. The use of an integrated form of ODE facilitates investigating the asymptotic behavior of the estimator and enhancing its robustness to the smoothing effect in the first step \citep{dattner2015optimal, chen2017network}. Both the two-step collocation methods proposed in this section involve maximizing the likelihood function for exponential family distributions, which can be efficiently solved with the iteratively reweighted least squares technique as in Section~\ref{sec:outer}. 

We compare the two-step collocation methods with the high-dimensional generalized profiling (HDGP) procedure in Section \ref{sec:simu}. For process and derivative estimation, since HDGP balances both the data and ODE fidelities, it usually results in reasonable fits and more accurate ODE parameter estimates due to the more accurate derivatives. For sparse structure identification, GRADE achieves the best accuracy, which is consistent with the motivation of GRADE for network reconstruction \citep{chen2017network}. In summary, HDGP is a better choice for process fitting and ODE parameter estimation, while GRADE excels in sparse structure identification.

\section{Simulation Studies}\label{sec:simu}
This section compares the empirical performance of three dynamical modeling approaches: the high-dimensional generalized profiling (HDGP) procedure and the two-step collocation methods proposed in Section~4, namely the GRADE and the vanilla two-step method, respectively.

Consider the ODE system studied by \cite{chen2017network} which consists of eight processes in four pairs, for $k=1,\ldots,4$,
\begin{equation*}
	\begin{cases}
		\theta'_{2k-1}(t) = 2k\pi\ \theta_{2k}(t)\\
		\theta'_{2k}(t) = 2k\pi\ \theta_{2k-1}(t)
	\end{cases}, \ t\in [0,1].
\end{equation*}
It is clear that the ODE solutions take the form of sine and cosine functions with varying frequencies, whereas no interaction exists across pairs. For the $k$th pair, the initial state is $\sin(y_k)$ and $\cos(y_k)$, where $y_k$ is sampled from $N(0,1)$.
The latent processes $\btheta(t)=(\theta_1(t),\dots,\theta_8(t))^\top$ described by the above ordinary differential equations are used to generate observations from Gaussian, Poisson and Bernoulli distributions. Denote by $t_1,\dots,t_n$ time points from $[0,1]$. For Gaussian distribution, $y_{ij}$ is sampled from $N(\theta_j(t_i),\sigma^2)$ with known variance $\sigma^2$, and the sample size $n$ for each process is set to be 100 and 500. For Poisson distribution, we draw 500 and 1000 samples from $\text{Poisson}(\lambda_j(t_i))$ where the intensity process $\lambda_j(t)=\exp\{\theta_j(t)\}$. For Bernoulli distribution, 1500 and 2500 samples are generated with probability of success $p_j(t)=\exp\{\theta_j(t)\}/[1+\exp\{\theta_j(t)\}]$. Sample sizes for Poisson and Bernoulli distributions are larger than Gaussian, as in those cases more observations are generally required to ensure reasonable estimates according to the theory of generalized linear model.

We use the smoothing spline fitting as an initialization for the profiling procedure, which also corresponds to the first stage of two-step collocation methods. The order of B-spline functions in HDGP is set as 6, and the number of knots is half of that of time points. Both HDGP and GRADE require numerical integration to evaluate ODE fidelity and integrated basis representations, respectively.
For sparsity penalty choices, we consider the Lasso penalty $\mathrm{PEN}_{\lambda_\gamma}(\bgamma_{j})=\lambda_\gamma \|\bgamma_j\|_1$ and the SCAD penalty $\textrm{PEN}_{\lambda_\gamma}(\bgamma_{j})=\sum_{k=1}^p p_{\lambda_\gamma}(|\gamma_{jk}|)$, where the function $p_\lambda(\cdot)$ is defined on $[0,\infty)$ as
$$
p_\lambda(u) = \left\{\begin{array}{ll}
	\lambda u, & \text{if $0\leq u \leq\lambda$} \\
	-(u^2 - 2a\lambda u + \lambda^2)/(a-1), & \text{if $\lambda< u <a\lambda$}\\
	(a+1)\lambda^2/2 & \text{if $ u \geq a\lambda$,}
\end{array}\right.
$$
and a suggested value for $a$ is 3.7 according to \cite{fan2001variable}. Algorithmic convergence is demonstrated when the difference between successive ODE parameter estimates is small enough. It works well for two-step collocation methods. However, due to the complex interaction between inner and outer optimizations, HDGP may not yield sparse ODE parameter estimates at the declaration of convergence. To address this numerical issue, we manually set ODE parameter estimates below a constant threshold as zero. Based on our empirical studies, a recommended value for the threshold is the root-mean-square of the initial estimate $\widehat\Gamma$ multiplied by a factor $0.01$.

Simulation results are evaluated using three types of criteria.
The first two criteria concern about process and derivative estimates, which are evaluated by the mean squared errors (MSE) of $\btheta(t)$ and $\btheta{'}(t)$,
\begin{align*}
\textrm{MSE}(\wh{\btheta}(t))=\frac{1}{np}\sum_{j=1}^p\sum_{i=1}^{n} \left\{ \wh\theta_j(t_i) - \theta_j(t_i)\right\}^2, \\
\textrm{MSE}(\wh{\btheta}'(t))=\frac{1}{np}\sum_{j=1}^p\sum_{i=1}^{n} \left\{ \wh\theta'_j(t_i) - \theta'_j(t_i)\right\}^2.
\end{align*}
Second, we measure how well the structural parameters are estimated by their root-mean-square error (RMSE). Third, true positive rate (TPR) and false positive rate (FPR) are used to quantify how well the sparse structure is identified, where we refer to non-zero structural parameters as positive cases and otherwise as negative cases. 

\begin{table}
	\centering
	\caption{Performance of HDGP, GRADE and the vanilla two-step method evaluated based on the process estimates ($\textrm{MSE}(\wh{\btheta}(t))$), derivative estimates ($\textrm{MSE}(\wh{\btheta}'(t))$), non-zero parameter estimation (RMSE), and sparse structure estimates (FPR). The 95\% confidence intervals are given in parentheses.}
	\scriptsize
	\begin{tabularx}{\textwidth}{lclRRRR}
		\toprule
		& N & Method & MSE ($\wh{\btheta}(t)$) & MSE ($\wh{\btheta}{'}(t)$) & RMSE ($\wh{\Gamma}$) & FPR \\
		\midrule
		\parbox[t]{2mm}{\multirow{6}{*}{\rotatebox[origin=c]{90}{Gaussian}}} & \multirow{3}{*}{100} & HDGP & \bigcell{r}{ 0.011 \\[-6pt] (0.0097,0.0124) } & \bigcell{r}{  \textbf{3.01} \\[-6pt] (2.48, 3.53) } & \bigcell{r}{ \textbf{0.58} \\[-6pt] (0.52,0.66) } & \bigcell{r}{ 0.44 \\[-6pt] (0.43,0.46) } \\
		&  & GRADE & \bigcell{r}{ \textbf{0.005} \\[-6pt] (0.0042,0.0049) } & \bigcell{r}{  5.23 \\[-6pt] (4.67, 5.87) } & \bigcell{r}{ 2.97 \\[-6pt] (2.89,3.05) } & \bigcell{r}{ \textbf{0.00} \\[-6pt] (-,-) } \\
		&  & vanilla & \bigcell{r}{ \textbf{0.005} \\[-6pt] (0.0043,0.0049) } & \bigcell{r}{  5.23 \\[-6pt] (4.74, 5.93) } & \bigcell{r}{ 0.62 \\[-6pt] (0.54,0.72) } & \bigcell{r}{ 0.89 \\[-6pt] (0.84,0.92) } \\
		\cmidrule(lr){2-7}
		
		& \multirow{3}{*}{500} & HDGP & \bigcell{r}{ 0.002 \\[-6pt] (0.0017,0.0023) } & \bigcell{r}{  \textbf{0.48} \\[-6pt] (0.41, 0.57) } & \bigcell{r}{ \textbf{0.28} \\[-6pt] (0.25,0.30) } & \bigcell{r}{ 0.44 \\[-6pt] (0.42,0.45) } \\
		&  & GRADE & \bigcell{r}{ \textbf{0.001} \\[-6pt] (0.0010,0.0011) } & \bigcell{r}{  1.82 \\[-6pt] (1.75, 1.88) } & \bigcell{r}{ 0.84 \\[-6pt] (0.82,0.85) } & \bigcell{r}{ \textbf{0.01} \\[-6pt] (0.01,0.02) } \\
		&  & vanilla & \bigcell{r}{ \textbf{0.001} \\[-6pt] (0.0010,0.0011) } & \bigcell{r}{  1.82 \\[-6pt] (1.75, 1.88) } & \bigcell{r}{ 0.34 \\[-6pt] (0.32,0.37) } & \bigcell{r}{ 0.67 \\[-6pt] (0.64,0.71) } \\
		\cmidrule(lr){1-7}
		
		\parbox[t]{2mm}{\multirow{6}{*}{\rotatebox[origin=c]{90}{Poisson}}} & \multirow{3}{*}{500} & HDGP & \bigcell{r}{ 0.024 \\[-6pt] (0.0222,0.0259) } & \bigcell{r}{  \textbf{6.24} \\[-6pt] (5.60, 6.93) } & \bigcell{r}{ \textbf{1.70} \\[-6pt] (1.54,1.91) } & \bigcell{r}{ 0.58 \\[-6pt] (0.56,0.61) } \\
		&  & GRADE & \bigcell{r}{ 0.024 \\[-6pt] (0.0232,0.0252) } & \bigcell{r}{ 12.27 \\[-6pt] (11.66,13.06) } & \bigcell{r}{ 2.03 \\[-6pt] (1.86,2.19) } & \bigcell{r}{ \textbf{0.37} \\[-6pt] (0.34,0.41) } \\
		&  & vanilla & \bigcell{r}{ 0.024 \\[-6pt] (0.0232,0.0252) } & \bigcell{r}{ 12.27 \\[-6pt] (11.63,13.00) } & \bigcell{r}{ 1.86 \\[-6pt] (1.70,2.07) } & \bigcell{r}{ 0.98 \\[-6pt] (0.97,0.98) } \\
		\cmidrule(lr){2-7}
		
		& \multirow{3}{*}{1000} & HDGP & \bigcell{r}{ \textbf{0.011} \\[-6pt] (0.0105,0.0121) } & \bigcell{r}{  \textbf{2.70} \\[-6pt] (2.43, 2.97) } & \bigcell{r}{ \textbf{1.04} \\[-6pt] (0.94,1.14) } & \bigcell{r}{ 0.57 \\[-6pt] (0.55,0.59) } \\
		&  & GRADE & \bigcell{r}{ 0.013 \\[-6pt] (0.0128,0.0141) } & \bigcell{r}{  8.20 \\[-6pt] (7.70, 8.73) } & \bigcell{r}{ 1.41 \\[-6pt] (1.31,1.53) } & \bigcell{r}{ \textbf{0.32} \\[-6pt] (0.28,0.36) } \\
		&  & vanilla & \bigcell{r}{ 0.013 \\[-6pt] (0.0127,0.0141) } & \bigcell{r}{  8.20 \\[-6pt] (7.71, 8.67) } & \bigcell{r}{ 1.18 \\[-6pt] (1.09,1.29) } & \bigcell{r}{ 0.97 \\[-6pt] (0.96,0.97) } \\
		\cmidrule(lr){1-7}
		
		\parbox[t]{2mm}{\multirow{6}{*}{\rotatebox[origin=c]{90}{Bernoulli}}} & \multirow{3}{*}{1500} & HDGP & \bigcell{r}{ 0.031 \\[-6pt] (0.0260,0.0357) } & \bigcell{r}{  \textbf{8.18} \\[-6pt] (6.74, 9.73) } & \bigcell{r}{ \textbf{1.77} \\[-6pt] (1.48,1.97) } & \bigcell{r}{ 0.54 \\[-6pt] (0.52,0.58) } \\
		&  & GRADE & \bigcell{r}{ 0.031 \\[-6pt] (0.0277,0.0333) } & \bigcell{r}{ 15.97 \\[-6pt] (14.77,16.99) } & \bigcell{r}{ 3.32 \\[-6pt] (3.01,3.61) } & \bigcell{r}{ \textbf{0.24} \\[-6pt] (0.21,0.28) } \\
		&  & vanilla & \bigcell{r}{ 0.032 \\[-6pt] (0.0285,0.0376) } & \bigcell{r}{ 22.28 \\[-6pt] (17.21,32.11) } & \bigcell{r}{ 2.28 \\[-6pt] (1.70,3.22) } & \bigcell{r}{ 0.94 \\[-6pt] (0.90,0.96) } \\
		\cmidrule(lr){2-7}
		
		& \multirow{3}{*}{2500} & HDGP & \bigcell{r}{ \textbf{0.019} \\[-6pt] (0.0169,0.0216) } & \bigcell{r}{  \textbf{5.05} \\[-6pt] (4.33, 6.03) } & \bigcell{r}{ \textbf{1.55} \\[-6pt] (1.39,1.74) } & \bigcell{r}{ 0.59 \\[-6pt] (0.55,0.62) } \\
		&  & GRADE & \bigcell{r}{ 0.020 \\[-6pt] (0.0193,0.0210) } & \bigcell{r}{ 12.71 \\[-6pt] (11.91,13.67) } & \bigcell{r}{ 2.57 \\[-6pt] (2.34,2.79) } & \bigcell{r}{ \textbf{0.20} \\[-6pt] (0.16,0.24) } \\
		&  & vanilla & \bigcell{r}{ 0.020 \\[-6pt] (0.0193,0.0211) } & \bigcell{r}{ 12.71 \\[-6pt] (11.88,13.69) } & \bigcell{r}{ 1.67 \\[-6pt] (1.46,1.87) } & \bigcell{r}{ 0.98 \\[-6pt] (0.97,0.99) } \\
		\bottomrule
	\end{tabularx}
	\label{tab:sim_compare1}
\end{table}
Table~\ref{tab:sim_compare1} displays the averaged evaluations over 50 repeated experiments using the Lasso penalty, while the true positive rates are omitted because they are all equal to one for all three methods. Under each simulation set-up, increasing the number of observations always leads to reduced errors and tighter confidence intervals in terms of the process fit and the parameter estimation. 
For process and derivative fitting, the smoothing splines method, as the first stage of two-step collocation methods, often produces accurate estimates of the latent process itself, but is less efficient in the derivative fitting. In contrast, the inner optimization of HDGP balances the data and ODE fidelities, resulting in reasonable process fitting and improved derivative fitting.
For ODE parameter estimation, HDGP delivers the smallest error due to the more accurate derivatives. Interestingly, GRADE has much worse performance than the other two under this criterion. One partial reason is that GRADE only uses structural parameters in the integrated basis representation (4.3) instead of the explicit form of differential equations.
For sparse structure identification, GRADE achieves the best accuracy, as it discovers all non-zero structural parameters with the fewest false positives. It is consistent with the motivation of GRADE for network reconstruction \citep{chen2017network}. In summary, HDGP is a better choice for process fitting and ODE parameter estimation, while GRADE excels in sparse structure identification.

\begin{figure}
	\centering
	\begin{subfigure}[b]{\textwidth}
		\includegraphics[width=\textwidth]{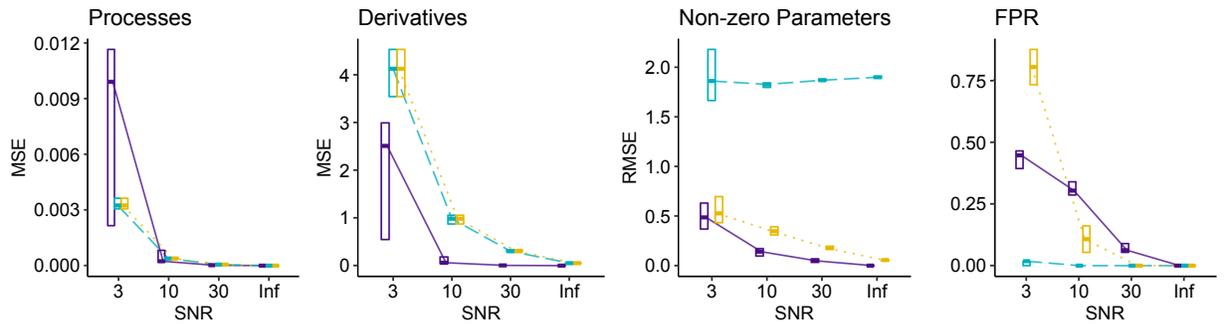}
		\caption{With Lasso penalty.}
		\label{fig:snr_lasso}
	\end{subfigure}
	
	\begin{subfigure}[b]{\textwidth}
		\includegraphics[width=\textwidth]{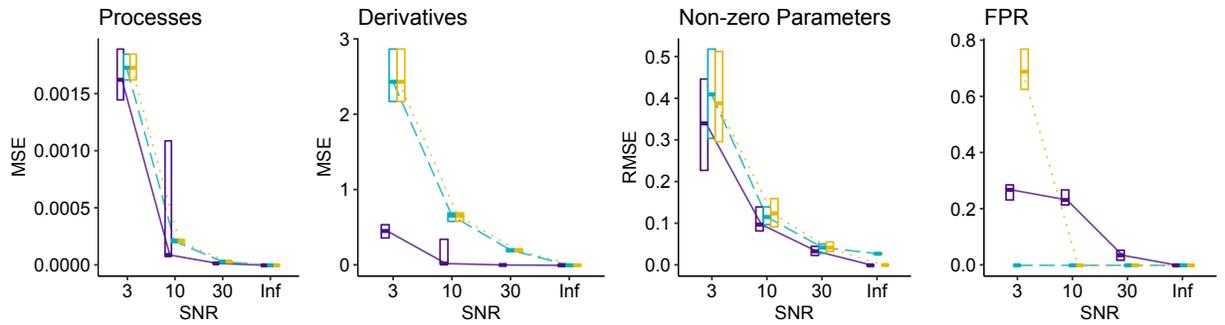}
		\caption{With SCAD penalty.}
		\label{fig:snr_scad}
	\end{subfigure}
	\caption{Performance of HDGP (purple solid), GRADE (blue dashed), and the vanilla two-step method (yellow dotted) for Gaussian observations at different noise levels. The boxes identify the medians and the quartiles of each criterion for 50 repeated experiments. Top and bottom rows correspond to Lasso and SCAD penalties, respectively.}
	\label{fig:snr}
\end{figure}

We next investigate the effects of different noise levels and choices of sparse penalty. Under the above Gaussian set-up with 500 observations for each process. The signal-to-noise ratio (SNR) is defined as the ratio between the sample standard deviation of $\{\theta_j(t_i)\}_{i=1}^n$ and the noise standard deviation $\sigma$. We set the signal-to-noise ratio as 3, 10, 30, and infinity, where the infinite ratio means that no noise is added. Both Lasso and SCAD penalties are considered. Figure~\ref{fig:snr} presents the performance evaluations over 50 repeated experiments. In general, all methods perform better over all criteria when the signal-to-noise ratio increases.
%
%
The top row of Figure~\ref{fig:snr} corresponding to the Lasso penalty provides the consistent result as in Table~\ref{tab:sim_compare1}, which indicates that HDGP has a comparable process fit and better derivative estimation, especially when the noise level is low.
Moreover, HDGP performs the best for estimating structural parameters, while the vanilla two-step method also provides satisfactory results. In contrast, even when there is no noise, the bias of ODE parameter estimates by GRADE is still large and RMSE is almost constant.
For sparse structure identification, GRADE outperforms the other methods under a wide range of noise levels. HDGP and the vanilla two-step method only have high accuracy when the signal level is high. 
The bottom row of Figure~\ref{fig:snr} displays simulation results when the SCAD penalty is used for inducing sparsity for the ODE system. Compared with the results with Lasso, overall performances in process, derivative, and ODE parameter estimations are improved mainly due to the unbiasedness property of SCAD penalty \citep{fan2001variable}. More interestingly, the poor performance of GRADE in ODE parameter estimation is greatly enhanced, and now it delivers comparable estimation results as the other two. Due to the oracle property enjoyed by the SCAD penalty \citep{fan2001variable}, we recommend it for better performance in parameter estimation.

\section{Real Data Analysis}\label{sec:app}

\subsection{Google Trends Data Analysis}

Google Trends provides a publicly accessible online portal to analyze the popularity of search queries. In this study, we attempt to apply our method to model the interactions among a number of trending keywords during the recent pandemic of Coronavirus disease 2019 (COVID-19). In Table \ref{tab:keyword}, we list 24 keywords and cluster them into three categories. The first category consists of five keywords about specific terminologies such as mask and quarantine. The second category includes not only the countries with the most confirmed cases as of January 2021, such as the United States, India, and Brazil but also the districts like Antarctica, which is the last continent to report confirmed cases due to the remoteness and sparse population. We also include the last category of noise keywords with no apparent relationship to the pandemic. 

\begin{table}[h!]
	\centering
	\caption{Three categories of keywords selected for the analysis of Google Trends data.}
	\small
	\begin{tabularx}{\textwidth}{@{}lX@{}}
		\toprule
		\textbf{Category} & \textbf{Keyword} \\
		\midrule
		COVID-19 related & coronavirus, mask, quarantine, vaccine, WHO (5 words)\\
		\midrule
		Countries or districts & Africa, Antarctica, Arctic, Australia, Brazil, Canada, China,
		India, Iran, Italy, Japan, Russia, the United States (13 words)\\
		\midrule
		Noise words & cat, cloud, desert, dog, game, sun (6 words) \\
		\bottomrule
	\end{tabularx}
	\label{tab:keyword}
\end{table}

\begin{figure}[h!]
	\centering
	\includegraphics[width=0.8\linewidth]{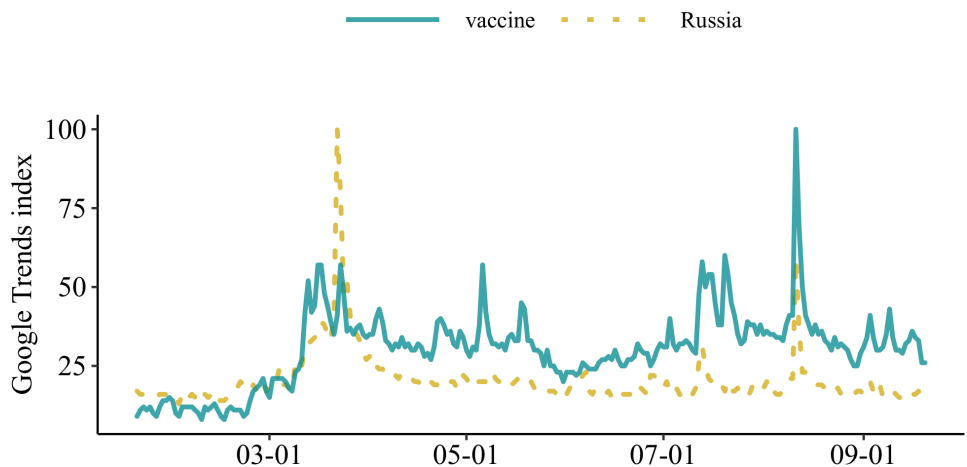}
	\caption{Daily Google Trends indices of keyword \textit{Russia} and \textit{vaccine} from January to September 2020.}
	\label{fig:gtrend_peaks}
\end{figure}

The Google Trends data used in our study cover the range from January 20 to September 20 in 2020. The keyword popularity is measured by an integer index calculated by normalizing and rounding the keyword count in an unbiased searching requests sample. We observe that the daily trend indices have several sharp peaks, see Figure \ref{fig:gtrend_peaks} for an illustrative example. Direct modeling for the mean trends will result in abrupt high values near the peaks and undersmooth other relatively flat regions. Therefore, it is more appropriate to assume the indices follow Poisson distributions, and we apply the proposed method to model the latent processes of intensity parameters with ODEs.

To better exhibit different stages of the pandemic, we consider three time periods: from January 20 to March 19, from March 20 to June 19, and from June 20 to September 20. For each period, our method is applied to fit the trending processes with a series of sparsity parameter $\lambda_\gamma$'s. 
\begin{figure}[h!]
	\centering
	\begin{subfigure}[b]{0.48\linewidth}
		\includegraphics[width=\textwidth]{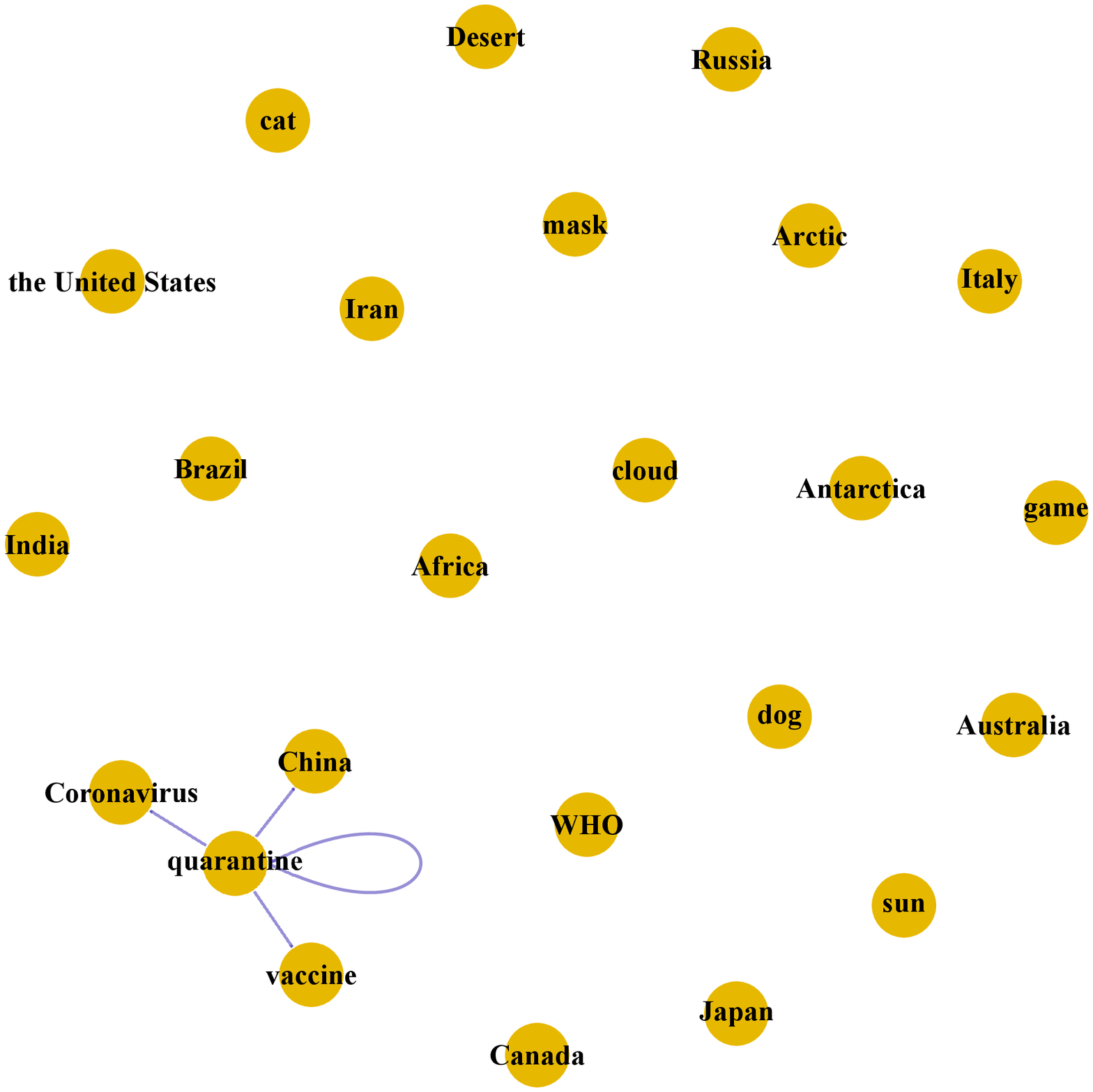}
		\caption{$\lambda_\gamma=10^{-0.5}$}
		\label{fig:covid_lam1}
	\end{subfigure}
	~
	\begin{subfigure}[b]{0.48\linewidth}
		\includegraphics[width=\textwidth]{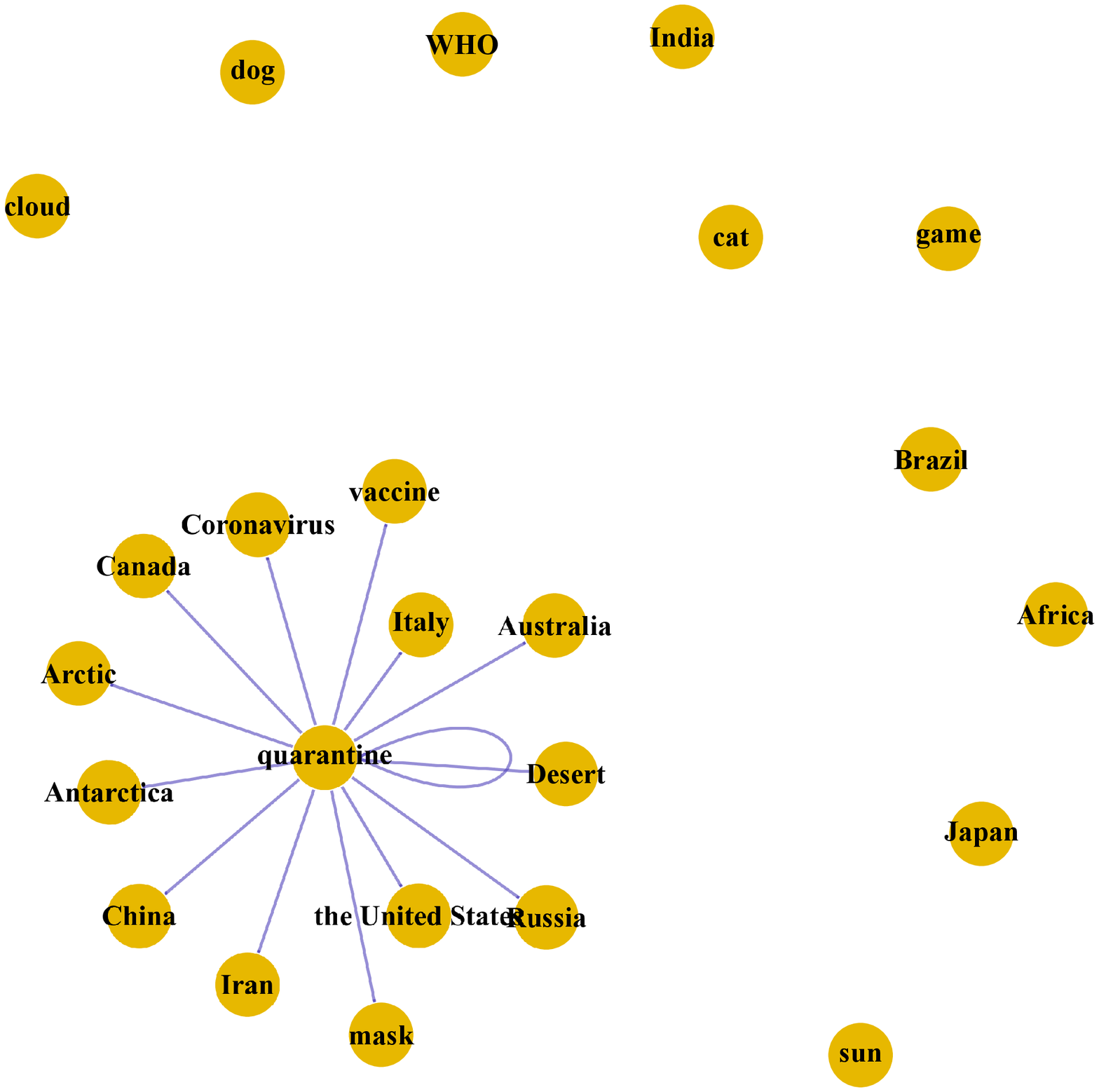}
		\caption{$\lambda_\gamma=10^{-1}$}
		\label{fig:covid_lam2}
	\end{subfigure}
	\caption{Recovered networks of the trending keywords during the first period (from January 20 to March 19) with different values of sparsity parameters.}
	\label{fig:covid_lam}
\end{figure}
Figures~\ref{fig:covid_lam1} and \ref{fig:covid_lam2} display two networks with different sparsity parameters in the first period (from January to March). Keyword \textit{quarantine} has the highest degree in both networks. During the COVID-19 pandemic, quarantines or self-quarantines are enacted by multiple governmental actors to prevent the rapid spread of the virus. It is of no surprise to become the top-ranked trending keyword. The other three keywords in Figure \ref{fig:covid_lam1} are \textit{coronavirus}, \textit{China} and \textit{vaccine}, which stand for the virus's name, the country where the first case was identified, and the immunization method. The top four keywords represent the major trending focus at the early stage of the pandemic. In Figure \ref{fig:covid_lam2}, more affected countries such as Australia, Italy, and the United States, are involved when the sparsity parameter is decreased. In contrast, noise keywords are isolated in both networks, indicating no connection to the trending topics.
\begin{table}[h!]
	\centering
	\caption{Top four keywords in the recovered networks during three periods. The keyword of the highest degree is in boldface.}
	\small
	\begin{tabularx}{\textwidth}{lX}
		\toprule
		Period  & Keywords \\
		\midrule
		January 20 -- March 19  & \textbf{quarantine}, China, coronavirus, vaccine \\
		March 20 -- June 19  & \textbf{Italy}, China, Iran, Russia \\
		June 20 -- September 20 & \textbf{coronavirus}, the United States, vaccine, mask\\
		\bottomrule
	\end{tabularx}
	\label{tab:covidnet}
\end{table}
More interestingly, we investigate the evolution of network structure for the trending keywords along the progression of the COVID-19 pandemic. Table \ref{tab:covidnet} lists the top four keywords in three time periods where the keyword with the highest degree is in boldface.
From the first period to the second, the keyword \textit{Italy} emerges as the new top word. 
According to the WHO report, on March 19, Italy overtook China as the country with the most reported deaths, and announced the national lockdown in March.
Turning to the third period, \textit{China} and \textit{Italy} drop out of the top list. Both countries had successfully slowed down the domestic infections and reduced daily new cases significantly. As preventive measures including wearing face masks in public are advised and several promising vaccines are being developed, \textit{mask} and \textit{vaccine} are among the top trending keywords.

\subsection{Analysis of Stock Price Change Directions}

In the year 2020, the stock market experienced enormous volatility due to the coronavirus pandemic. Many companies have suffered massive price drops, while others have witnessed substantial increases. We collect the stock price indices for 40 companies during 251 trading days spanning from January 1 to December 30, 2020. Our goal in this study is to characterize the change direction patterns of stock prices, taking into account the dynamic interactions among the stocks. To this end, the original price indices are coded as binary data to denote an increase or decrease. We group the companies into eight categories based on the Global Industry Classification Standard. Details are provided in Table~\ref{tab:stocks}. 
\begin{table}
	\small
	\centering
	\caption{Companies selected in eight categories for stock price data analysis.}
	\begin{tabularx}{\textwidth}{@{}clX@{}}
		\toprule
		Group & \textbf{Category} & \textbf{Companies}  \\
		\midrule
		1 & Information Technology & Adobe, Apple, Microsoft, Salesforce, Zoom \\
		2 & Electric Vehicle & BYD, Kandi, Nio, Tesla, Workhorse  \\
		3 & Pharmaceutical & AbbVie, Eli lilly, Moderna, Novartis, Pfizer \\
		4 & Consumer Services \& Retail & Ascena, J. C. Penney, Kohl's, Macy's, Nordstrom \\
		5 & Online Retail Shopping & Amazon, Best Buy, Target, Walmart, Wayfair \\
		6 & Hotels & Hilton, Marriott, Wyndham, Wynn, Park \\
		7 & Air Transportation & Boeing, Airbus, Delta Air Lines, Southwest Airlines, United Airline  \\
		8 & Energy & Chevron, Conocophillips, Exxon Mobil, Schlumberger, Valero Energy \\
		\bottomrule
	\end{tabularx}
	\label{tab:stocks}
\end{table}

\begin{figure}
	\centering
	\includegraphics[width=\textwidth]{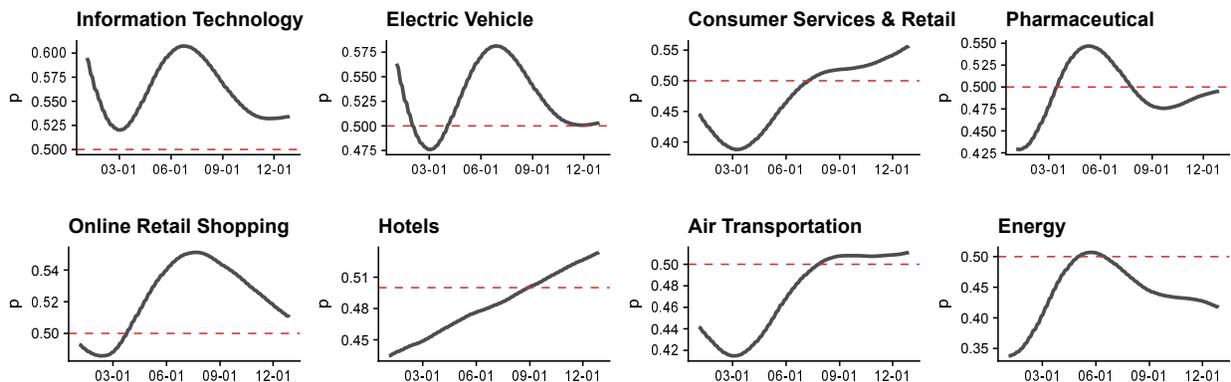}
	\caption{The fitted probability processes of daily price increase for the eight categories. The red dashed line denotes $p=0.5$.}
	\label{fig:stock_fit}
\end{figure}

The high-dimensional ODE system is built up for the latent success probability processes. Our sparsity tuning procedure leads to $\lambda_\gamma=10^{-2.1}$ and the fitted model achieves an ODE fidelity below $10^{-6}$.
Figure \ref{fig:stock_fit} displays the fitted probabilities of a daily stock price increase for all categories. 
We notice some interesting results from the result. First, all categories have the low fitted probabilities around March. It corresponds to the 2020 stock market crash, during which multiple circuit breakers were triggered on fears of the COVID-19 coronavirus. Since the crash, some sectors recovered and re-entered a bull market through December. Online retail companies made huge profits as health concerns changed customers' shopping habits. Information technology companies benefited from the growing demands for information services and electronics devices. For example, the shifts towards remote working had raised the number of Zoom's daily users to an unprecedented one.
In contrast, sectors like energy, hotels, and air transportation
experienced the most severe hit by the COVID-19 pandemic. Although there were signs of recovery in the fourth quarter, these industries are still under the tremendous impact of the COVID-19 recession.

\subsection{Analysis of Yeast Cell Cycle-regulated Genes}

The cell cycle is a fundamental biological process consisting of cell growth, duplication of genetic information, distribution of chromosomes, and cell division \citep{cho1998genome}.
\citet{spellman1998comprehensive} analyzed the expression levels of 6,178 yeast genes at 7-minute intervals for 119 minutes. The experiments were carried out in the cell cultures with three independent synchronization methods. A score was calculated for each gene to indicate their similarities to those cell-cycle regulated genes already known. 
Due to missingness in data, we choose 72 out of 93 genes identified by \citet{spellman1998comprehensive} in the $alpha$ factor-based synchronized experiment, and model the dynamic relationship between the mean profiles of these 72 genes using an ODE system under Gaussian assumption for gene expression level. The proposed method is applied to identify the sparse structure of the gene regulatory network. 
\begin{figure}
	\centering
	\includegraphics[width=0.65\textwidth]{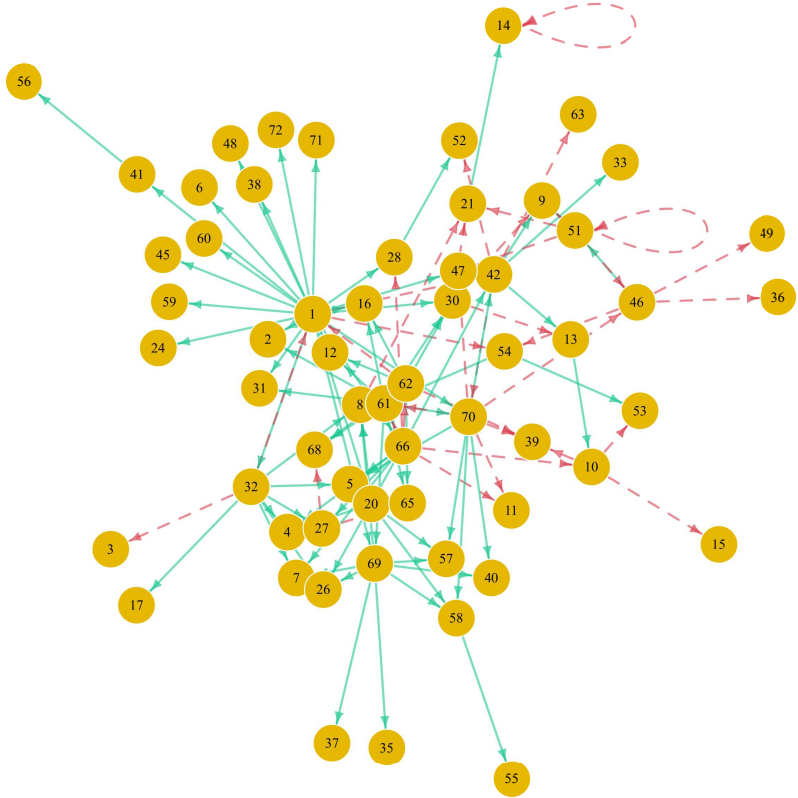}
	\caption{The recovered network of the yeast cell cycle. Yellow nodes represent genes, and the green-solid or red-dashed edges indicate potential promotion or suppression effects.}
	\label{fig:yeast}
\end{figure}
The result is shown in  Figure~\ref{fig:yeast}, which excludes 12 isolated genes. This suggests that although those genes get involved in the cell cycle, their regulated transcriptions are not absolutely required. 
Among the 60 genes in Figure~\ref{fig:yeast}, 116 regulations (i.e., directed edges) are discovered. The average number of regulations for each gene is around three, while more than 80\% genes have regulations fewer than five. Genes with high network degrees are identified as central hub nodes. For example, CLN3 (node 1) has the largest number of regulations in Figure~\ref{fig:yeast}. According to the Saccharomyces Genome Database \citep{cherry1998sgd}, the encoded protein CLN3p is known as a cell cycle regulator and promotes the G1/S transition \citep{nasmyth1993control}. More interestingly, the positive or negative signs of our estimated structural parameters naturally imply the potential promotion or inhibition between genes, respectively. Our result suggests that CHS1 (node 62) promotes the expression of POL30 (node 12), which regulates DNA replication in the G1 phase. Meanwhile, it suppresses the expression of FAR1 (node 30), which is a CDC28p kinase inhibitor functioning in the G2/M transition.

\section{Conclusions and Discussion}\label{sec:conclude}

In this article, we have proposed a new profiling procedure for both parameter estimation and sparse structure identification for high-dimensional linear ODE models with non-Gaussian observations. Our method involves a hierarchical optimization scheme: the inner optimization balances the data fitting and ODE fidelity to improve estimation efficiency, while the outer optimization induces a sparse structure for better model interpretation. Besides, we extend two-step collocation methods to the non-Gaussian observation setting and compare them with the proposed profiling procedure via comprehensive studies.

One limitation of our work is that only the linear ODE system is under consideration. We are aware of the recent development of two-step collocation to sparse additive ODE systems \citep{henderson2014network, wu2014sparse,chen2017network} and a more general functional ANOVA extension \citep{dai2021kernel}. Although our hierarchical optimization is not restricted to the linear ODE, the extension to nonlinear ODE systems is not straightforward. For instance, a common strategy to handle additive ODE models is to expand the nonlinear components with basis function. However, due to the profiling nature, the range of collocation bases for latent processes needs to be controlled within a compact interval, which may not be easily overcome. Another future research is on the statistical properties such as uniform bound on the approximations to the true solutions, asymptotic normality of the estimators. Despite existing theory established for the standard generalized profiling \citep{qi2010asymptotic}, it is still a challenging problem due to high dimensionality, and we leave the systematic study to future work.



\appendix

\section{Derivatives}\label{sec:derivative}
We provide the analytical expressions of the derivatives used in the computation (Section 3).

\subsection*{Derivatives of $G_j$ in inner optimization} \label{sec:appendix_deriv_Gj}

Write $G_j(\theta_j; \bgamma_j)$ in the inner optimization 
\begin{align*}
	G_j = -\frac{1}{n} \sum_{i=1}^n \left\{ y_{ij} \theta_j(t_i) - b(\theta_j(t_i)) \right\} +
	\lambda_{\theta} \int_\mathcal{T} \left\{ \theta'_j(t) - \gamma_{j0} - 
	\sum_{k=1}^{p} \gamma_{jk}\, \theta_k(t) \right\}^2 \,\mathrm{d}t,
\end{align*}
then the first derivative is 
\begin{align*} 
	\frac{\partial G_j}{\partial \bc_j}
	=& -\frac{1}{n}\sum_{i=1}^n\{y_{ij}\bh(t_i)-b'(\theta_j(t_i)) \bh(t_i)\}\\
	&\qquad + 2\lambda_\theta \int_\mathcal{T} 
	\biggl\{ 
	\frac{\mathrm{d}\theta_j(t)}{\mathrm{d}t} - \gamma_{j0} - \sum_{k=1}^p \gamma_{jk}\, \theta_k(t)
	\biggr\}
	\biggl\{ 
	\frac{\mathrm{d}\bh(t)}{\mathrm{d}t} - \gamma_{jj} \, \bh(t)
	\biggr\}
	\,\mathrm{d}t,
\end{align*}
and the second derivative is
\begin{align*} 
	\frac{\partial^2 G_j}{\partial \bc_j \partial \bc_j^\top}
	=& \frac{1}{n}\sum_{i=1}^n\{b''(\theta_j(t_i)) \bh(t_i)\bh(t_i)^\top\}\\
	&\qquad+ 2\lambda_\theta \int_\mathcal{T} 
	\biggl\{ 
	\frac{\mathrm{d}\bh(t)}{\mathrm{d}t} - \gamma_{jj}  \bh(t)
	\biggr\} \biggl\{ 
	\frac{\mathrm{d}\bh(t)}{\mathrm{d}t} - \gamma_{jj}  \bh(t)
	\biggr\}^\top 
	\,\mathrm{d}t.
\end{align*}
For $k=0$,
\begin{align*}
	\frac{\partial^2 G_j}{\partial \bc_j \partial \gamma_{j0}}
	&= -2\lambda_\theta \int_\mathcal{T} 
	\biggl\{ 
	\frac{\mathrm{d}\bh(t)}{\mathrm{d}t} - \gamma_{jj} \bh(t)
	\biggr\}
	\,\mathrm{d}t,
\end{align*}
for $k=1,\dots,p$ and $k\neq j$ ,
\begin{align*} 
	\frac{\partial^2 G_j}{\partial \bc_j \partial \gamma_{jk}}
	&= -2\lambda_\theta \int_\mathcal{T} 
	\biggl\{ 
	\frac{\mathrm{d}\bh(t)}{\mathrm{d}t} - \gamma_{jj}  \bh(t)
	\biggr\}
	\theta_k(t)
	\,\mathrm{d}t,
\end{align*}
for $k=j$,
\begin{align*}
	\frac{\partial^2 G_j}{\partial \bc_j \partial \gamma_{jj}}
	&= -2\lambda_\theta \int_\mathcal{T} 
	\biggl\{ 
	\frac{\mathrm{d}\bh(t)}{\mathrm{d}t} - \gamma_{jj} \bh(t)
	\biggr\}
	\theta_j(t)
	\,\mathrm{d}t \\
	&\qquad -2\lambda_\theta \int_\mathcal{T} 
	\biggl\{ 
	\frac{\mathrm{d}\theta_j(t)}{\mathrm{d}t} - \gamma_{j0} - \sum_{k=1}^p \gamma_{jk} \theta_k(t)
	\biggr\}
	\bh(t)
	\,\mathrm{d}t.
\end{align*}

\subsection*{Derivative of $\bc_j^*$ in outer optimization}

Write $\bc_j^*(\bgamma_j)$ as $\bc_j^*$ for simplicity.
Since $G_j$ has zero-gradient at $\bc_j^*$, then
$$
\frac{\partial G_j}{\partial \bc_j} \bigg|_{\bc_j^*} =0.
$$
Taking the derivative with respect to $\bgamma_j$ on both sides, we have
$$
\frac{\mathrm{d}}{\mathrm{d}\bgamma_j^\top}\biggl(\frac{\partial G_j}{\partial \bc_j}\bigg|_{\bc_j^*}\biggr)
=\frac{\partial^2 G_j}{\partial\bc_j \partial\bgamma_j^\top}\bigg|_{\bc_j^*} 
+ \left( \frac{\partial^2 G_j}{\partial\bc_j \partial\bc_j ^\top}\bigg|_{\bc_j^*}\right) 
\frac{\partial \bc_j^*(\bgamma_j)}{\partial\bgamma_j^\top} 
=0.
$$
Suppose $\partial^2 G_j/(\partial\bc_j \partial\bc_j ^\top)|_{\bc_j^*}$ is non-singular, we have the following expression of the derivative
\begin{equation*}
	\frac{\partial \bc_j^*(\bgamma_j)}{\partial\bgamma_j^\top} 
	=-\biggl(\frac{\partial^2 G_j}{\partial\bc_j \partial\bc_j ^\top}\bigg|_{\bc_j^*}\biggr)^{-1}
	\biggl(\frac{\partial^2 G_j}{\partial\bc_j \partial\bgamma_j^\top}\bigg|_{\bc_j^*} \biggr).
\end{equation*}
Both matrices on the right-hand side have been explicitly derived, the the derivative of $\bc_j^*$ follows.

\bigskip
\begin{center}
{\large\bf SUPPLEMENTARY MATERIALS}
\end{center}

\begin{description}

\item[Supplementary Material] contains additional numerical results.

\end{description}

\bibliographystyle{Chicago}

\bibliography{bib-ode-exp.bib}

\begin{thebibliography}{}

\bibitem[\protect\citeauthoryear{Brunel, Clairon, and d’Alch{\'e} Buc}{Brunel
  et~al.}{2014}]{brunel2014parametric}
Brunel, N.~J., Q.~Clairon, and F.~d’Alch{\'e} Buc (2014).
\newblock Parametric estimation of ordinary differential equations with
  orthogonality conditions.
\newblock {\em Journal of the American Statistical Association\/}~{\em
  109\/}(505), 173--185.

\bibitem[\protect\citeauthoryear{Cao and Ramsay}{Cao and
  Ramsay}{2007}]{cao2007parameter}
Cao, J. and J.~O. Ramsay (2007).
\newblock Parameter cascades and profiling in functional data analysis.
\newblock {\em Computational Statistics\/}~{\em 22\/}(3), 335--351.

\bibitem[\protect\citeauthoryear{Carey and Ramsay}{Carey and
  Ramsay}{2021}]{carey2021fast}
Carey, M. and J.~O. Ramsay (2021).
\newblock Fast stable parameter estimation for linear dynamical systems.
\newblock {\em Computational Statistics \& Data Analysis\/}~{\em 156}, 107124.

\bibitem[\protect\citeauthoryear{Chen, Shojaie, and Witten}{Chen
  et~al.}{2017}]{chen2017network}
Chen, S., A.~Shojaie, and D.~M. Witten (2017).
\newblock Network reconstruction from high-dimensional ordinary differential
  equations.
\newblock {\em Journal of the American Statistical Association\/}~{\em
  112\/}(520), 1697--1707.

\bibitem[\protect\citeauthoryear{Cherry, Adler, Ball, Chervitz, Dwight, Hester,
  Jia, Juvik, Roe, and Schroeder}{Cherry et~al.}{1998}]{cherry1998sgd}
Cherry, J.~M., C.~Adler, C.~Ball, S.~A. Chervitz, S.~S. Dwight, E.~T. Hester,
  Y.~Jia, G.~Juvik, T.~Roe, and M.~Schroeder (1998).
\newblock Sgd: {S}accharomyces genome database.
\newblock {\em Nucleic Acids Research\/}~{\em 26\/}(1), 73--79.

\bibitem[\protect\citeauthoryear{Cho, Campbell, Winzeler, Steinmetz, Conway,
  Wodicka, Wolfsberg, Gabrielian, Landsman, and Lockhart}{Cho
  et~al.}{1998}]{cho1998genome}
Cho, R.~J., M.~J. Campbell, E.~A. Winzeler, L.~Steinmetz, A.~Conway,
  L.~Wodicka, T.~G. Wolfsberg, A.~E. Gabrielian, D.~Landsman, and D.~J.
  Lockhart (1998).
\newblock A genome-wide transcriptional analysis of the mitotic cell cycle.
\newblock {\em Molecular cell\/}~{\em 2\/}(1), 65--73.

\bibitem[\protect\citeauthoryear{Cokus, Feng, Zhang, Chen, Merriman,
  Haudenschild, Pradhan, Nelson, Pellegrini, and Jacobsen}{Cokus
  et~al.}{2008}]{cokus2008shotgun}
Cokus, S.~J., S.~Feng, X.~Zhang, Z.~Chen, B.~Merriman, C.~D. Haudenschild,
  S.~Pradhan, S.~F. Nelson, M.~Pellegrini, and S.~E. Jacobsen (2008).
\newblock Shotgun bisulphite sequencing of the arabidopsis genome reveals dna
  methylation patterning.
\newblock {\em Nature\/}~{\em 452\/}(7184), 215--219.

\bibitem[\protect\citeauthoryear{Dai and Li}{Dai and Li}{2021}]{dai2021kernel}
Dai, X. and L.~Li (2021).
\newblock Kernel ordinary differential equations.
\newblock {\em Journal of the American Statistical Association\/}.

\bibitem[\protect\citeauthoryear{Dattner and Klaassen}{Dattner and
  Klaassen}{2015}]{dattner2015optimal}
Dattner, I. and C.~A. Klaassen (2015).
\newblock Optimal rate of direct estimators in systems of ordinary differential
  equations linear in functions of the parameters.
\newblock {\em Electronic Journal of Statistics\/}~{\em 9\/}(2), 1939--1973.

\bibitem[\protect\citeauthoryear{Dodds, Harris, Kloumann, Bliss, and
  Danforth}{Dodds et~al.}{2011}]{dodds2011temporal}
Dodds, P.~S., K.~D. Harris, I.~M. Kloumann, C.~A. Bliss, and C.~M. Danforth
  (2011).
\newblock Temporal patterns of happiness and information in a global social
  network: Hedonometrics and twitter.
\newblock {\em PLoS ONE\/}~{\em 6\/}(12), e26752.

\bibitem[\protect\citeauthoryear{Fan, Feng, and Wu}{Fan et~al.}{2009}]{fan2009}
Fan, J., Y.~Feng, and Y.~Wu (2009, 06).
\newblock Network exploration via the adaptive lasso and scad penalties.
\newblock {\em The Annals of Applied Statistics\/}~{\em 3\/}(2), 521--541.

\bibitem[\protect\citeauthoryear{Fan and Li}{Fan and
  Li}{2001}]{fan2001variable}
Fan, J. and R.~Li (2001).
\newblock Variable selection via nonconcave penalized likelihood and its oracle
  properties.
\newblock {\em Journal of the American statistical Association\/}~{\em
  96\/}(456), 1348--1360.

\bibitem[\protect\citeauthoryear{Fan, Li, Zhang, and Zou}{Fan
  et~al.}{2020}]{fan2020statistical}
Fan, J., R.~Li, C.-H. Zhang, and H.~Zou (2020).
\newblock {\em Statistical foundations of data science}.
\newblock Chapman and Hall/CRC.

\bibitem[\protect\citeauthoryear{Gu}{Gu}{2013}]{gu2013smoothing}
Gu, C. (2013).
\newblock {\em Smoothing Spline ANOVA Models\/} (2nd ed.), Volume 297.
\newblock New York: Springer.

\bibitem[\protect\citeauthoryear{Hall and Ma}{Hall and
  Ma}{2014}]{hall2014quick}
Hall, P. and Y.~Ma (2014).
\newblock Quick and easy one-step parameter estimation in differential
  equations.
\newblock {\em Journal of the Royal Statistical Society: Series B (Statistical
  Methodology)\/}~{\em 76\/}(4), 735--748.

\bibitem[\protect\citeauthoryear{Hecker, Lambeck, Toepfer, Van~Someren, and
  Guthke}{Hecker et~al.}{2009}]{hecker2009gene}
Hecker, M., S.~Lambeck, S.~Toepfer, E.~Van~Someren, and R.~Guthke (2009).
\newblock Gene regulatory network inference: data integration in dynamic
  models—a review.
\newblock {\em Biosystems\/}~{\em 96\/}(1), 86--103.

\bibitem[\protect\citeauthoryear{Henderson and Michailidis}{Henderson and
  Michailidis}{2014}]{henderson2014network}
Henderson, J. and G.~Michailidis (2014).
\newblock Network reconstruction using nonparametric additive {ODE} models.
\newblock {\em PLoS ONE\/}~{\em 9\/}(4), e94003.

\bibitem[\protect\citeauthoryear{Huang, Nakamori, and Wang}{Huang
  et~al.}{2005}]{huang2005forecasting}
Huang, W., Y.~Nakamori, and S.-Y. Wang (2005).
\newblock Forecasting stock market movement direction with support vector
  machine.
\newblock {\em Computers \& operations research\/}~{\em 32\/}(10), 2513--2522.

\bibitem[\protect\citeauthoryear{Liang and Wu}{Liang and
  Wu}{2008}]{liang2008parameter}
Liang, H. and H.~Wu (2008).
\newblock Parameter estimation for differential equation models using a
  framework of measurement error in regression models.
\newblock {\em Journal of the American Statistical Association\/}~{\em
  103\/}(484), 1570--1583.

\bibitem[\protect\citeauthoryear{Lu, Liang, Li, and Wu}{Lu
  et~al.}{2011}]{lu2011high}
Lu, T., H.~Liang, H.~Li, and H.~Wu (2011).
\newblock High-dimensional odes coupled with mixed-effects modeling techniques
  for dynamic gene regulatory network identification.
\newblock {\em Journal of the American Statistical Association\/}~{\em
  106\/}(496), 1242--1258.

\bibitem[\protect\citeauthoryear{Ma, Zhang, Huang, and Zhong}{Ma
  et~al.}{2017}]{ma2017adaptive}
Ma, P., N.~Zhang, J.~Z. Huang, and W.~Zhong (2017).
\newblock Adaptive basis selection for exponential family smoothing splines
  with application in joint modeling of multiple sequencing samples.
\newblock {\em Statistica Sinica\/}~{\em 27\/}(4), 1757--1777.

\bibitem[\protect\citeauthoryear{McCullagh and Nelder}{McCullagh and
  Nelder}{1989}]{mccullagh1989generalized}
McCullagh, P. and J.~Nelder (1989).
\newblock {\em Generalized Linear Models, Second Edition}.
\newblock Chapman \& Hall/CRC Monographs on Statistics \& Applied Probability.
  Taylor \& Francis.

\bibitem[\protect\citeauthoryear{Miao, Wu, and Xue}{Miao
  et~al.}{2014}]{miao2014generalized}
Miao, H., H.~Wu, and H.~Xue (2014).
\newblock Generalized ordinary differential equation models.
\newblock {\em Journal of the American Statistical Association\/}~{\em
  109\/}(508), 1672--1682.

\bibitem[\protect\citeauthoryear{Nagalakshmi, Wang, Waern, Shou, Raha,
  Gerstein, and Snyder}{Nagalakshmi
  et~al.}{2008}]{nagalakshmi2008transcriptional}
Nagalakshmi, U., Z.~Wang, K.~Waern, C.~Shou, D.~Raha, M.~Gerstein, and
  M.~Snyder (2008).
\newblock The transcriptional landscape of the yeast genome defined by {RNA}
  sequencing.
\newblock {\em Science\/}~{\em 320\/}(5881), 1344--1349.

\bibitem[\protect\citeauthoryear{Nasmyth}{Nasmyth}{1993}]{nasmyth1993control}
Nasmyth, K. (1993).
\newblock Control of the yeast cell cycle by the {C}dc28 protein kinase.
\newblock {\em Current Opinion in Cell Biology\/}~{\em 5\/}(2), 166--179.

\bibitem[\protect\citeauthoryear{Polynikis, Hogan, and di~Bernardo}{Polynikis
  et~al.}{2009}]{polynikis2009comparing}
Polynikis, A., S.~Hogan, and M.~di~Bernardo (2009).
\newblock Comparing different {ODE} modelling approaches for gene regulatory
  networks.
\newblock {\em Journal of Theoretical Biology\/}~{\em 261\/}(4), 511--530.

\bibitem[\protect\citeauthoryear{Powell}{Powell}{2006}]{powell2006newuoa}
Powell, M.~J. (2006).
\newblock The {NEWUOA} software for unconstrained optimization without
  derivatives.
\newblock In {\em Large-scale nonlinear optimization}, pp.\  255--297.
  Springer.

\bibitem[\protect\citeauthoryear{Poyton, Varziri, McAuley, McLellan, and
  Ramsay}{Poyton et~al.}{2006}]{poyton2006parameter}
Poyton, A., M.~S. Varziri, K.~B. McAuley, P.~J. McLellan, and J.~O. Ramsay
  (2006).
\newblock Parameter estimation in continuous-time dynamic models using
  principal differential analysis.
\newblock {\em Computers \& Chemical Engineering\/}~{\em 30\/}(4), 698--708.

\bibitem[\protect\citeauthoryear{Qi and Zhao}{Qi and
  Zhao}{2010}]{qi2010asymptotic}
Qi, X. and H.~Zhao (2010).
\newblock Asymptotic efficiency and finite-sample properties of the generalized
  profiling estimation of parameters in ordinary differential equations.
\newblock {\em The Annals of Statistics\/}~{\em 38\/}(1), 435--481.

\bibitem[\protect\citeauthoryear{Ramsay}{Ramsay}{1996}]{ramsay1996principal}
Ramsay, J.~O. (1996).
\newblock Principal differential analysis: Data reduction by differential
  operators.
\newblock {\em Journal of the Royal Statistical Society: Series B (Statistical
  Methodology)\/}~{\em 58\/}(3), 495--508.

\bibitem[\protect\citeauthoryear{Ramsay, Hooker, Campbell, and Cao}{Ramsay
  et~al.}{2007}]{ramsay2007parameter}
Ramsay, J.~O., G.~Hooker, D.~Campbell, and J.~Cao (2007).
\newblock Parameter estimation for differential equations: a generalized
  smoothing approach.
\newblock {\em Journal of the Royal Statistical Society: Series B (Statistical
  Methodology)\/}~{\em 69\/}(5), 741--796.

\bibitem[\protect\citeauthoryear{Sloan and Morgan}{Sloan and
  Morgan}{2015}]{sloan2015tweets}
Sloan, L. and J.~Morgan (2015).
\newblock Who tweets with their location? understanding the relationship
  between demographic characteristics and the use of geoservices and geotagging
  on twitter.
\newblock {\em PLoS ONE\/}~{\em 10\/}(11), e0142209.

\bibitem[\protect\citeauthoryear{Spellman, Sherlock, Zhang, Iyer, Anders,
  Eisen, Brown, Botstein, and Futcher}{Spellman
  et~al.}{1998}]{spellman1998comprehensive}
Spellman, P.~T., G.~Sherlock, M.~Q. Zhang, V.~R. Iyer, K.~Anders, M.~B. Eisen,
  P.~O. Brown, D.~Botstein, and B.~Futcher (1998).
\newblock Comprehensive identification of cell cycle-regulated genes of the
  yeast {S}accharomyces cerevisiae by microarray hybridization.
\newblock {\em Molecular Biology of the Cell\/}~{\em 9\/}(12), 3273--3297.

\bibitem[\protect\citeauthoryear{Stuart, Segal, Koller, and Kim}{Stuart
  et~al.}{2003}]{stuart2003gene}
Stuart, J.~M., E.~Segal, D.~Koller, and S.~K. Kim (2003).
\newblock A gene-coexpression network for global discovery of conserved genetic
  modules.
\newblock {\em Science\/}~{\em 302\/}(5643), 249--255.

\bibitem[\protect\citeauthoryear{Tibshirani}{Tibshirani}{1996}]{tibshirani1996regression}
Tibshirani, R. (1996).
\newblock Regression shrinkage and selection via the lasso.
\newblock {\em Journal of the Royal Statistical Society: Series B (Statistical
  Methodology)\/}~{\em 58\/}(1), 267--288.

\bibitem[\protect\citeauthoryear{Tseng and Yun}{Tseng and
  Yun}{2009}]{tseng2009coordinate}
Tseng, P. and S.~Yun (2009).
\newblock A coordinate gradient descent method for nonsmooth separable
  minimization.
\newblock {\em Mathematical Programming\/}~{\em 117\/}(1), 387--423.

\bibitem[\protect\citeauthoryear{Varah}{Varah}{1982}]{varah1982spline}
Varah, J.~M. (1982).
\newblock A spline least squares method for numerical parameter estimation in
  differential equations.
\newblock {\em SIAM Journal on Scientific and Statistical Computing\/}~{\em
  3\/}(1), 28--46.

\bibitem[\protect\citeauthoryear{Voorman, Shojaie, and Witten}{Voorman
  et~al.}{2014}]{voorman2014graph}
Voorman, A., A.~Shojaie, and D.~Witten (2014).
\newblock Graph estimation with joint additive models.
\newblock {\em Biometrika\/}~{\em 101\/}(1), 85--101.

\bibitem[\protect\citeauthoryear{Wahba, Wang, Gu, Klein, and Klein}{Wahba
  et~al.}{1995}]{wahba1995}
Wahba, G., Y.~Wang, C.~Gu, R.~Klein, and B.~Klein (1995, 12).
\newblock Smoothing spline {ANOVA} for exponential families, with application
  to the {Wisconsin} epidemiological study of diabetic retinopathy : the 1994
  {Neyman} memorial lecture.
\newblock {\em The Annals of Statistics\/}~{\em 23\/}(6), 1865--1895.

\bibitem[\protect\citeauthoryear{Wang and Leng}{Wang and
  Leng}{2008}]{wang2008note}
Wang, H. and C.~Leng (2008).
\newblock A note on adaptive group lasso.
\newblock {\em Computational Statistics \& Data Analysis\/}~{\em 52\/}(12),
  5277--5286.

\bibitem[\protect\citeauthoryear{Wood}{Wood}{2017}]{wood2017generalized}
Wood, S.~N. (2017).
\newblock {\em Generalized additive models: an introduction with R}.
\newblock CRC press.

\bibitem[\protect\citeauthoryear{Wu, Lu, Xue, and Liang}{Wu
  et~al.}{2014}]{wu2014sparse}
Wu, H., T.~Lu, H.~Xue, and H.~Liang (2014).
\newblock Sparse additive ordinary differential equations for dynamic gene
  regulatory network modeling.
\newblock {\em Journal of the American Statistical Association\/}~{\em
  109\/}(506), 700--716.

\bibitem[\protect\citeauthoryear{Wu, Qiu, Yuan, and Wu}{Wu
  et~al.}{2019}]{wu2019parameter}
Wu, L., X.~Qiu, Y.-x. Yuan, and H.~Wu (2019).
\newblock Parameter estimation and variable selection for big systems of linear
  ordinary differential equations: a matrix-based approach.
\newblock {\em Journal of the American Statistical Association\/}~{\em
  114\/}(526), 657--667.

\bibitem[\protect\citeauthoryear{Yuan and Kendziorski}{Yuan and
  Kendziorski}{2006}]{yuan2006hidden}
Yuan, M. and C.~Kendziorski (2006).
\newblock Hidden {Markov} models for microarray time course data in multiple
  biological conditions.
\newblock {\em Journal of the American Statistical Association\/}~{\em
  101\/}(476), 1323--1332.

\bibitem[\protect\citeauthoryear{Yuan and Lin}{Yuan and
  Lin}{2006}]{yuan2006model}
Yuan, M. and Y.~Lin (2006).
\newblock Model selection and estimation in regression with grouped variables.
\newblock {\em Journal of the Royal Statistical Society: Series B (Statistical
  Methodology)\/}~{\em 68\/}(1), 49--67.

\bibitem[\protect\citeauthoryear{Yuan and Lin}{Yuan and
  Lin}{2007}]{yuan2007model}
Yuan, M. and Y.~Lin (2007).
\newblock Model selection and estimation in the {Gaussian} graphical model.
\newblock {\em Biometrika\/}~{\em 94\/}(1), 19--35.

\bibitem[\protect\citeauthoryear{Zhang}{Zhang}{2010}]{zhang2010nearly}
Zhang, C.-H. (2010).
\newblock Nearly unbiased variable selection under minimax concave penalty.
\newblock {\em The Annals of Statistics\/}~{\em 38\/}(2), 894--942.

\bibitem[\protect\citeauthoryear{Zhang, Conn, and Scheinberg}{Zhang
  et~al.}{2010}]{zhang2010derivative}
Zhang, H., A.~R. Conn, and K.~Scheinberg (2010).
\newblock A derivative-free algorithm for least-squares minimization.
\newblock {\em SIAM Journal on Optimization\/}~{\em 20\/}(6), 3555--3576.

\bibitem[\protect\citeauthoryear{Zou}{Zou}{2006}]{zou2006adaptive}
Zou, H. (2006).
\newblock The adaptive lasso and its oracle properties.
\newblock {\em Journal of the American Statistical Association\/}~{\em
  101\/}(476), 1418--1429.

\end{thebibliography}
\end{document}